\documentclass[useAMS,usenatbib]{mn2e}

\usepackage{graphicx}
\usepackage{natbib}
\usepackage{longtable}
\usepackage[T1]{fontenc}
\usepackage{aecompl}

\newcommand{\iras}{\textit{IRAS}~19135$+$3937}
\newcommand{\bd}{BD$+$46$\degr$442}

\title[\iras: An SRd binary with a disc]{\iras: An SRd variable as interacting binary surrounded by a circumbinary disc.}
\author[N. Gorlova et al.]{N.~Gorlova,$^{1}$\thanks{E-mail: n27032001@yahoo.com}
H.~Van~Winckel,$^{1}$ N.~P.~Ikonnikova,$^{2}$ M.~A.~Burlak,$^{2}$ G.~V.~Komissarova,$^{2}$
\newauthor 
A.~Jorissen,$^{3}$ C.~Gielen,$^{4}$ J.~Debosscher,$^{1}$ P.~Degroote$^{1}$
\\
$^{1}$Instituut voor Sterrenkunde, Katholieke Universiteit Leuven,
Celestijnenlaan 200D, 3001, Leuven, Belgium\\
$^{2}$Lomonosov Moscow State University, Sternberg Astronomical
Institute, 13 Universitetskij prospekt, 119234, Moscow, Russia\\
$^{3}$Institut d'Astronomie et d'Astrophysique, Universit\'{e} Libre de Bruxelles,
CP 226, Boulevard du Triomphe, 1050, Bruxelles, Belgium\\
$^{4}$Belgian Institute for Space Aeronomy, 1180, Brussels, Belgium\\}

\begin{document}

\date{Accepted Received}

\pagerange{\pageref{firstpage}--\pageref{lastpage}} \pubyear{2002}

\maketitle

\label{firstpage}

\begin{abstract}

Semi-regular (SR) variables are not a homogeneous class and their
variability is often explained due to pulsations and/or binarity.
This study focuses on \iras, an SRd variable
with an infra-red excess indicative of a dusty disc.
A time-series of high-resolution spectra, $UBV$ photometry as well as
a very accurate light curve obtained by the {\it Kepler} satellite, allowed
us to study the object in unprecedented detail.
We discovered it to be a binary with a period of 127 days.
The primary has a low surface gravity and an atmosphere
depleted in refractory elements.
This combination of properties unambiguously
places \iras\, in the subclass of post-Asymptotic Giant Branch stars with dusty discs.

We show that the light variations in this object can not be due to
pulsations, but are likely caused by the obscuration of the primary by
the circumbinary disc during orbital motion.  Furthermore, we argue
that the double-peaked Fe emission lines provide evidence for the
existence of a gaseous circumbinary Keplerian disc inside the dusty
disc.  A secondary set of absorption lines has been detected near
light minimum, which we attribute to the reflected spectrum of the
primary on the disc wall, which segregates due to the different
Doppler shift. This corroborates the recent finding that reflection in
the optical by this type of discs is very efficient.  The system also
shows a variable H$\alpha$ profile indicating a collimated outflow
originating around the companion.
\iras\, thus encompasses all the major emergent trends about evolved
disc systems, that will eventually help to place these objects in the
evolutionary context.

\end{abstract}


\section{Introduction}

There are four types of semi-regular (SR) variables, each designated alphabetically
from a to d. \iras\, belongs to the last category, "d".
SRd variables represent a poorly understood
group of warm (spectral types F--K) giants and supergiants.
Their light curves exhibit variations on the time-scales 30--1100 days, that are only
quasi-regular and lack characteristic features of the radial pulsators
with similar periods, such as Cepheids and RV Tau stars.  There
is, however, one intriguing feature in the behaviour of some RV Tau
stars that is reminiscent of the SR variability.  Normally RV Tau
stars pulsate with periods between 30--150 days, with two alternating
minima per cycle, shallow and deep.  An 'RV Tau b' subgroup, however,
shows an additional variability in the mean magnitude, amplitude, or
relative strengths of the minima.
This additional 'secondary period' variability occurs on the time scale
that overlaps with variability in the SRd-s with long periods. This phenomenon
is referred to as 'long secondary periods' (LSP), and the origin of LSPs is
still a matter of debate \citep[e.g.,][]{Kiss2007,Wood1999}. The two major hypotheses discussed in
the literature are variable obscuration \citep{LloydEvans1974,Fokin1994,Pollard1997} and an
interplay between various pulsation modes \citep{Buchler1987}.

Modelling of the better studied Galactic systems
is hindered by the lack of direct distance measurements, hence luminosities.
In terms of kinematics, chemical composition, and emission line strength
there appears to be no systematic difference
between SRd-s and RV Tau stars \citep{Wahlgren1993}.
Chemical composition studies indicate that both groups are heterogeneous and
include halo, as well as disc objects \citep{Giridhar2000, Britavskiy2012}.
Based on the rarity, presence of the circumstellar matter,
and the luminosities deduced for the Magellanic Clouds objects,
most of these objects must represent late evolutionary stages of 
low mass stars, being either post-red giant or post-asymptotic giant branch (pAGB)
stars \citep{Wallerstein2002,VanWinckel2003, Kamath2014}.
Since there is no evidence that SRd-s and RV Tau stars would represent
different stellar populations or different evolutionary stages,
the reason for the semi-regular behaviour must be sought
in the properties of individual systems.

Binarity could be such a property, but how exactly would it cause the
observed light variations? Mutual eclipses of the companions can be
ruled out, as the fraction of systems with an edge-on orientation or
where both companions are evolved giants, must be
negligible. Furthermore, the semi-regular character requires that the
eclipsing body should be variable.  It has long been known that many
RV Tau stars have near-infrared excesses, indicating presence of hot
dust up to the sublimation temperatures \citep{Evans1985}.  The
modelling of the spectral energy distribution (SED)  \citep{Gielen2009}, as
well as interferometric measurements \citep{Deroo2007, Hillen2013}
have demonstrated that the dust resides in a disc.  The discs have typical inner
radii of only a few astronomical units ($\sim$10 stellar radii,
R$_{\star}$) and inner walls of substantial scale-height.

At the same time, there is a growing evidence from the radial velocity
(RV) studies that evolved objects with discs are all binaries
\citep{VanWinckel1999,VanWinckel2009}.  Based on the deduced orbits
for the visible giant primaries, the discs must be circumbinary.  The
variability can then be naturally explained by the obscuration of the
primary star by the inner rim of the disc, as the line of sight to the
star probes different heights above the disc plane along its orbit
\citep{Waelkens1993}.  The cycle-to-cycle irregularities could be
caused by inhomogeneity or precession of the disc.

To test this theory, one has to find a correlation between the
semi-regular behaviour, the shape of the SED, and binarity.  In
reality, it is not always possible to prove binarity based on RV
variations alone, for example in case of small inclination angle or
strong pulsations.  In 2009 we started to monitor a number of RV Tau,
SR, W Vir, and chemically peculiar pAGB stars with the echelle
spectrograph HERMES \citep{VanWinckel2010, Gorlova2011}.  Besides
detecting RV variations consistent with binarity and obtaining orbital
solutions for some systems, in many of them we also discovered
specific phase-dependent H$\alpha$ profiles
\citep{Gorlova2012a,VanWinckel2012,Gorlova2013}.  Normally these
profiles are P Cyg-like or have a double-peak emission, but only
during the superior conjunction of the pAGB primary they were found to
develop strong blue-shifted absorption.  \citet{Witt2009}, based on
the observations of a similar phenomenon in the central star of the
Red Rectangle (RR) nebula, explained the H$\alpha$ line-profile
variability by a model in which a jet is powered by
accretion from the giant primary to the invisible (likely a
main-sequence, MS) companion.  \iras, originally included in our
sample due to the disc-like infra-red (IR) excess, turned out to be
one of such objects. Besides H$\alpha$, it caught our eye also because
of the rather smooth light-curve with a large amplitude, which is not
typical for long-period pulsators.  \iras\, thus presented a good case
for testing binary theory for semi-regular variables.

The variability of \iras\, has been first discovered by amateur
astronomers \citep{Sallman2004}, who determined a period $P$=125.4 d
and an amplitude $\Delta V=0.9$ mag.  Based on these observations the
star was included in the General Catalogue of Variable Stars as an SRd
variable V677 Lyr \citep{Kazarovets2013}.  The star was also observed
in the course of the All Sky Automated Survey (ASAS), where it was
classified as a ``QPER'' type, which is a semi-regular variable with a
stable period (128.8 d, \citealt{Pigulski2009}).  The star is
relatively faint ($V\sim11$ mag) and had not been discussed in the
literature when we started its observations. Recently, \citet{Rao2014}
performed an abundance analysis and found it to be a moderately
metal-poor star ([Fe/H]=$-$1 dex) with some peculiarities.

In \citet{Gorlova2012b} we presented the first RV curve of \iras\,
that revealed its binarity.
We also showed a number of peculiar spectral lines including H$\alpha$,
and pointed to the similarity with \bd,
the first object discovered in our survey to display this type of
H$\alpha$ line profile variability \citep{Gorlova2012a}.
Since then we have tripled the number of observed spectra of \iras\, and followed up with
multi-band photometry.
Here, we provide a comprehensive analysis of these data
in order to understand the cause of the semi-regular variability in
this object. We deduce strong observational constraints on the
geometry of the system and study the ongoing interaction processes.

\section{New observations and data reduction}

\subsection{Photometry}

We observed \iras\, with a photoelectric $UBV$-photometer
\citep{Lyutyj1971} attached to the 60-cm telescope Zeiss-1 (located in Crimea)
of the Sternberg Astronomical Institute (SAI Russia). The diameter of the photometer
diaphragm was set to 27 arcsec. Seventy four measurements have been
made between 2012 -- 2013, which overlaps with the epoch of our spectroscopic observations.

To obtain colour terms for the transformation of the instrumental $ubv$ system
to the standard Johnson's $UBV$ one, we observed standard stars in NGC 6633 \citep{Mermilliod1986}.
By solving equations $\Delta V = \Delta v - k_{1}\Delta (b-v) $, $\Delta (B-V) = k_{2}\Delta(b-v) $,
and $\Delta (U-B) = k_{3}\Delta (u-b)$, we obtained the following colour terms:
$k_{1}=0.082\pm0.010$, $k_{2}=0.980\pm0.010$, $k_{3}=1.081\pm0.015$.
We used HD 179909 as a comparison star,
with the following magnitudes adopted from Mermilliod (1991)\footnote{VizieR on-line catalogue II/168 ``Homogeneous Means in the UBV System''.}:
$V=8.285 \pm 0.015$, $B-V=-0.070 \pm 0.010$,
$U-B=-0.330 \pm 0.010$.
The typical accuracy of our absolute photometry is of the order of 0.01 mag.

The measurements are given in Table \ref{tab_ubv}.
The observations revealed sinusoidal-like variations in brightness and colour,
with a period consistent with earlier reports.

\subsection{Spectroscopy}

We collected 61 spectra of \iras\, with the HERMES fibre echelle spectrograph mounted
on the 1.2 m telescope Mercator on La Palma \citep{Raskin2011}.
The observations were carried out in the years 2009 -- 2013,
with an average cadence of one observation per two weeks.
The exposure time varied with the brightness of the star and the weather conditions,
being on average 1400 s. A typical signal-to-noise ratio (S/N) in the central orders (around H$\alpha$) was $\sim$35,
which was needed for obtaining a cross-correlation function
suitable for the RV determination (Section \ref{sec_rv}). 
All spectra in our survey have been collected using the high-resolution fibre configuration,
that provides $R\sim 80,000$ and a wavelength coverage between 3800
and 9000 \AA,
and reduced with a dedicated \textsc{Python}-based pipeline \textsc{HermesDRS}.

The pipeline provides science graded output.
A quick-look of the extracted spectrum is available during the actual observations,
and the full reduction is automatically performed after sunrise.
The pipeline first averages in 2D biases, arcs, and flat fields.
The cross-order profile with the two slices is modelled on a daily basis
and used at extraction, during which the cosmic clipping is also performed. 
We do not use optimal extraction. The read-out noise is typically 3--4 electrons.
Sky subtraction is not performed, because all spatial information is
lost due to the scrambling properties of the optical fibre.
The contamination by the interstellar emission, however,
is not a concern for our evolved stars,
that are located away from star-forming regions. 

The stability of the zero-point of our arc-based wavelength calibration is
monitored with the help of the International Astronomical Union RV standards.
Over 5 years of HERMES operation we obtain a standard deviation of 0.080 km/s
based on 2329 measurements of different radial-velocity standards,
measured as a spread over the mean of every standard. The shifts are mainly cased
by the pressure variations during the night \citep{Raskin2011}.
More details on the instrument and data reduction can be found
at the HERMES webpage\footnote{http://www.mercator.iac.es/instruments/hermes/} and in \citet{Gorlova2012a}.

\section{Basic properties}

\subsection{Photospheric parameters and chemical composition}\label{chem_comp}

For the chemical abundance study we used an average of three well exposed consecutive spectra
obtained on 29 June 2009, with maximum S/N$\sim$150 near 6000 \AA\, in the combined spectrum.
These observations were carried out near maximum light,
and metallic lines were nearly symmetric and easy to measure.
As will be shown in Sec. \ref{sec_rv}, this phase corresponds
to the superior conjunction of the pAGB primary, when its obscuration by
the circumstellar disc and the contribution from a putative companion should be minimal.

We followed the same procedure for the physical parameters and abundance determination
as for another similar object \bd, as described in \citet{Gorlova2011phys} and \citet{Gorlova2012a}. 
Briefly, to get the initial estimate of the effective temperature ($T_{\mathrm{eff}}$)
and surface gravity ($\log g$), we matched the observed profiles
of H$\delta$, H$\gamma$, and H$\beta$ (wings only) with the precomputed state of the art model profiles
of \citet{Coelho2005}, and Paschen 14 with those of \citet{Munari2000}.
The two distinct sets of model spectra had to be employed in order to cover both Balmer and Paschen series.
After visual inspection, we selected the following best combinations
of $T_{\mathrm{eff}}$/$\log g$
\footnote{The following units are used throughout the paper: K for the effective temperature $T_{\mathrm{eff}}$,
dex[cm s$^{-2}$] for the logarithm of surface gravity $\log g$, km s$^{-1}$ for the micro-turbulent velocity $V_{\mathrm{tur}}$, and dex for the logarithm of the elemental abundance with respect to the hydrogen
abundance (using designation (X/H) for log $\mathrm{\epsilon (X)}$
on the scale where log $\mathrm{\epsilon (H)}=12$,
and [X/H] when expressing difference with the Sun: [X/H]$=$(X/H)$-$(X/H)$_{\odot}$).}:
6000/1.0$\pm$0.5, 6250/2.0, and 5750/0.5.

At the next stage of iteration, we examined the correlation of the
Fe abundance obtained from the Fe\,\textsc{i} and Fe\,\textsc{ii} lines with their equivalent widths (EWs).
The abundances (for Fe, as well as other elements)
were computed from the EWs using \textsc{MOOG10}, which is the latest
version of the LTE abundance determination code by C. Sneden \citep{Sneden1973},
and the ATLAS9 model atmospheres of R. Kurucz in the updated version of \citet{Castelli2003}.
We started with models with the solar metallicity, but re-computed the final abundances
with $\mathrm{[M/H]}=-1.0$ to better match the deduced overall metal deficiency of \iras.
The EWs in the observed spectrum were measured using our \textsc{Python}-based program,
which includes a module to disentangle close line pairs. The latter is important for \iras,
because the lines are relatively broad (FWHM = 12.5 km~s$^{\mathrm{-1}}$ in the considered phase).
To maximize the number of measured lines, we also augmented our previous atomic line list
(which was an up-dated version from \citealt{Kovtyukh1999}), with two others: a list previously used by the Leuven group
\citep{VanWinckel2000}, which is largely based on the solar list by F. Th\'{e}venin \citep{Thevenin1989,Thevenin1990},
and a \textsc{SpectroWeb}\footnote{http://spectra.freeshell.org/spectroweb.html} list,
as compiled and improved by A. Lobel \citep{Lobel2008, Lobel2011}.     
Where the oscillator strengths (log $gf$) for the same line differed between
the lists by more than 0.25 dex, the value was used
that provided an abundance best matching the rest of the lines of the same ion.

\begin{figure}
 \includegraphics[width=84mm]{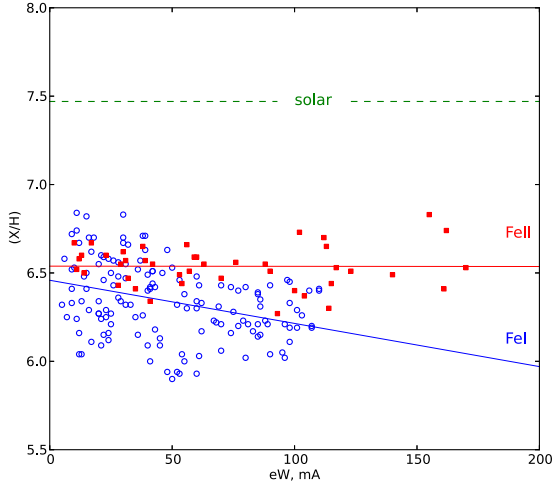}
 \caption{Ionization balance for Fe lines, that was used to derive the following atmospheric parameters for \iras:
 $V_{\mathrm{tur}}$=5 km~s$^{\mathrm{-1}}$, $\log g$=1.0, log $\mathrm{\epsilon (Fe) =6.50}$. The plot is shown for $T_{\mathrm{eff}}$=6000 K, which
 is the best-fit value from the fit to the Hydrogen lines. Straight lines designate linear fits to the data.}
 \label{fig_Fe_EW}
\end{figure}

To study the Fe\,\textsc{i}/Fe\,\textsc{ii} balance,
the Fe abundance was computed with \textsc{MOOG10} for three best values of $T_{\mathrm{eff}}$ (5750, 6000, and 6250 K),
a range of $\log g=$0.5--3.0 with a step of 0.5 dex, and a range of micro-turbulences 
$V_{\mathrm{tur}}$=3--8 km~s$^{\mathrm{-1}}$ with a step of 1 km~s$^{\mathrm{-1}}$. Only Fe\,\textsc{i} lines with EW$\le$110 m\AA\,
and Fe\,\textsc{ii} lines with EW$\le$200 m\AA\, were retained for further analysis, as 
they lie on the linear part of the curve of growth.
For each model, the derived abundances from the individual lines have been plotted against the EWs. 
We obtained different values for
the micro-turbulence when trying to remove the abundance trends with EWs
for the neutral and ionic transitions.
As discussed in \citet{Gorlova2012a}, at these temperatures and gravities
strong Fe\,\textsc{i} lines are susceptible to deviations from local thermodynamic equilibrium (LTE),
therefore, $V_{\mathrm{tur}}$ from the Fe\,\textsc{ii} lines was adopted.
The surface gravity is then established by requiring that the
average Fe abundance from the Fe\,\textsc{ii} lines matches the
extrapolated to EW=0 abundance from the Fe\,\textsc{i} lines (Fig. \ref{fig_Fe_EW}).
Following this procedure for each of the three values of $T_{\mathrm{eff}}$ obtained from the
hydrogen line analysis, we obtained the following  $T_{\mathrm{eff}}$/$\log g$/$V_{\mathrm{tur}}$/[Fe/H] solutions:
6000/1.0/5.0/$-$0.97 (best model), 6250/1.5/6.0/$-$0.84, and 5750/0.5/5.0/$-$1.12,
where the Fe abundance is expressed relative to the solar value of 7.47
in the scale log $\mathrm{\epsilon (H)=12}$.
 
Similar rules were used for the rest of the elements 
to combine abundances from the individual lines: for ions
the abundances of all lines with EW $\le$ 200 m\AA\, were averaged;
for neutrals an extrapolation to EW=0 was performed for lines with EW $\le$ 110 m\AA\,
when possible, otherwise, abundances of lines with EW $\le$ 50 m\AA\, were averaged,
or, if such lines did not exist (Co\,\textsc{i}, Zn\,\textsc{i}), lines with EW up to 110 m\AA\, have
been averaged. These final abundances for the best atmospheric model are given in Table \ref{Tab_abund},
along with the difference with abundances obtained using two other closest models.

\begin{figure}
 \includegraphics[width=84mm]{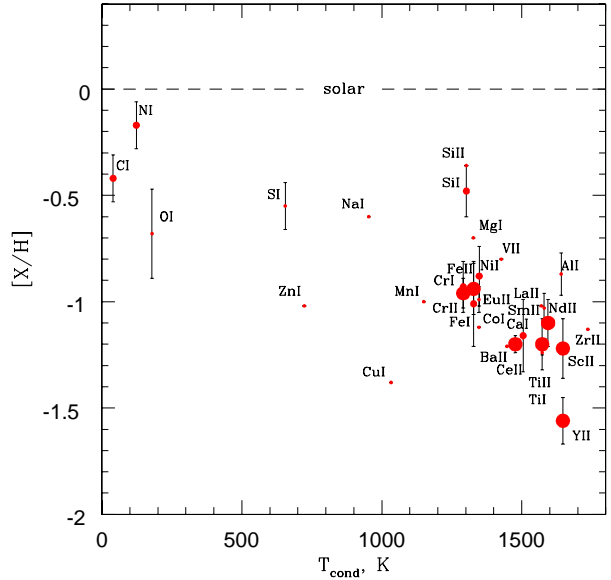}
 \caption{Abundances of \iras\, (computed with the best-fit model $T_{\mathrm{eff}}$/$\log g$/$V_{\mathrm{tur}}$=6000/1.0/5.0)
 versus condensation temperature.
The errorbars represent the line-to-line scatter, while the circle size designates
the reliability flag: the larger the circle, the more reliable the flag.}
 \label{fig_ab_trend}
\end{figure}

All elements in \iras\, show sub-solar abundance, ranging from $-$0.5 for the CNO group
to $-$1.5 for some heavy elements. The under-abundance appears to correlate with the condensation temperature
of the element, as shown in Fig. \ref{fig_ab_trend}.
This is a common phenomenon in pAGB disc systems.
It is explained by contamination of the photosphere by the re-processed gas from the disc.
The gas is poor in refractory elements because they condensed into grains.
\citet{Rao2014} carried an independent analysis of the \iras\, composition,
in the framework of their survey of objects occupying the RV Tau box on the \textit{IRAS} colour-colour diagram.
The spectrum that they investigated was obtained in the phase of line doubling,
and so is less favourable for measuring EWs than ours.
Nevertheless, they obtained very similar atmospheric parameters including metallicity:
$T_{\mathrm{eff}}$/$\log g$/$V_{\mathrm{tur}}$/[Fe/H]$=$6000/0.5/4.1/$-$1.04.
The depletion pattern that we find for this star is relatively weak,
so it is not surprising that \citet{Rao2014} did not manage to detect it.
They did point out the deficiency of the $\alpha$-elements Ca and Ti,
that have a higher condensation temperature than Fe, but were reluctant to ascribe it
to the depletion effect due to a lack of a stronger Sc deficiency and of the Zn enrichment.  
Our values for these elements are more consistent with the depletion pattern,
except for Zn, but the latter is only represented by 1 or 2 lines in both studies.

\begin{table}
\caption{Chemical composition of \iras}\label{Tab_abund}
\begin{tabular}{@{}rcrrccccc}
 \hline
 $Z$ & Ion & log $\epsilon$ & $\rmn{[X/H]}$ & rms & \multicolumn{2}{c}{$\Delta \rmn{[X/H]}$}  & $N$ & Flag\\
\cline{6-7} 
     &     &                &       &     &  $A$         & $B$ &            &         \\  
 \hline
 6 & C\,\textsc{i} & 8.10 & $-$0.42 & 0.11 & $-$0.03 & $+$0.03 & 8 & 2 \\
 7 & N\,\textsc{i} & 7.76 & $-$0.17 & 0.11 & $+$0.03 & $-$0.05 & 4 & 2 \\
 8 & O\,\textsc{i} & 8.21 & $-$0.68 & 0.21 & $-$0.09 & $+$0.09 & 2 & 3 \\
 11 & Na\,\textsc{i} & 5.72 & $-$0.60 & $-$ & $-$0.08 & $+$0.08 & 1 & 3 \\
 12 & Mg\,\textsc{i} & 6.88 & $-$0.70 & $-$ & $-$0.05 & $+$0.03 & 1 & 3 \\
 13 & Al\,\textsc{i} & 5.60 & $-$0.87 & 0.10 & $-$0.05 & $+$0.04 & 2 & 3 \\
 14 & Si\,\textsc{i} & 7.06 & $-$0.48 & 0.12 & $-$0.07 & $+$0.07 & 13 & 2 \\
 14 & Si\,\textsc{ii} & 7.26 & $-$0.36 & $-$ & $-$0.01 & $-$0.09 & 1 & 3 \\
 16 & S\,\textsc{i} & 6.57 & $-$0.55 & 0.11 & $-$0.09 & $+$0.09 & 2 & 3 \\
 20 & Ca\,\textsc{i} & 5.20 & $-$1.16 & 0.17 & $-$0.11 & $+$0.09 & 12 & 2 \\
 21 & Sc\,\textsc{ii} & 1.93 & $-$1.22 & 0.14 & $-$0.26 & $+$0.22 & 11 & 1 \\
 22 & Ti\,\textsc{i} & 3.77 & $-$1.24 & 0.01 & $-$0.21 & $+$0.19 & 2 & 3 \\
 22 & Ti\,\textsc{ii} & 3.81 & $-$1.20 & 0.12 & $-$0.24 & $+$0.19 & 24 & 1 \\
 23 & V\,\textsc{ii} & 3.17 & $-$0.80 & $-$ & $-$0.22 & $+$0.19 & 1 & 3 \\
 24 & Cr\,\textsc{i} & 4.68 & $-$0.93 & 0.12 & $-$0.11 & $+$0.11 & 8 & 2 \\
 24 & Cr\,\textsc{ii} & 4.63 & $-$0.96 & 0.07 & $-$0.16 & $+$0.13 & 13 & 1 \\
 25 & Mn\,\textsc{i} & 4.38 & $-$1.0 & $-$ & $-$0.14 & $+$0.13 & 1 & 3 \\
 26 & Fe\,\textsc{i} & 6.46 & $-$1.01 & 0.20 & $-$0.13 & $+$0.13 & 142 & 2 \\
 26 & Fe\,\textsc{ii} & 6.54 & $-$0.94 & 0.12 & $-$0.18 & $+$0.14 & 43 & 1 \\
 27 & Co\,\textsc{i} & 3.77 & $-$1.12 & $-$ & $-$0.12 & $+$0.12 & 1 & 3 \\
 28 & Ni\,\textsc{i} & 5.35 & $-$0.88 & 0.14 & $-$0.13 & $+$0.14 & 27 & 2 \\
 29 & Cu\,\textsc{i} & 2.83 & $-$1.38 & $-$ & $-$0.2 & $+$0.2 & 1 & 3 \\
 30 & Zn\,\textsc{i} & 3.55 & $-$1.02 & $-$ & $-$0.15 & $+$0.15 & 1 & 3 \\
 39 & Y\,\textsc{ii} & 0.66 & $-$1.56 & 0.11 & $-$0.26 & $+$0.25 & 6 & 1 \\
 40 & Zr\,\textsc{ii} & 1.45 & $-$1.13 & $-$ & $-$0.24 & $+$0.25 & 1 & 3 \\
 56 & Ba\,\textsc{ii} & 0.92 & $-$1.21 & $-$ & $-$0.3 & $+$0.26 & 1 & 3 \\
 57 & La\,\textsc{ii} & 0.17 & $-$1.02 & $-$ & $-$0.3 & $+$0.3 & 1 & 3 \\
 58 & Ce\,\textsc{ii} & 0.33 & $-$1.20 & 0.04 & $-$0.29 & $+$0.29 & 3 & 1  \\
 60 & Nd\,\textsc{ii} & 0.46 & $-$1.10 & 0.11 & $-$0.31 & $+$0.32 & 4 & 1 \\
 62 & Sm\,\textsc{ii} & $-$0.04 & $-$1.03 & 0.07 & $-$0.31 & $+$0.32 & 2 & 3 \\
 63 & Eu\,\textsc{ii} & $-$0.49 & $-$0.99 & 0.06 & $-$0.26 & $+$0.27 & 2 & 3 \\
 \hline
\end{tabular}

\medskip
Column [X/H] gives abundances for the best-fit model $T_{\mathrm{eff}}$/$\log g$/$V_{\mathrm{tur}}$=6000/1.0/5.0,
while the $\Delta \rmn{[X/H]}$ column shows the response of the abundances to the change
in the adopted $T_{\mathrm{eff}}$, with cases A and B corresponding to the models 5750/0.5/5.0
and 6250/1.5/6.0.
The flags in the last column have been introduced as follows:
1 -- most reliable abundances, for ions where three and more lines with EW$\le$200 m\AA\, were available for
averaging; 2 -- less reliable abundances, obtained for neutrals by extrapolating to EW=0
the abundances from lines with EW$\le$110 m\AA, or if not possible, by averaging lines with EW$\le$50 m\AA;
3 -- least reliable abundances, obtained by averaging less than three lines with EW$\le$200 mA and EW$\le$110 m\AA\, for ions and neutrals, respectively.
The number of used lines is indicated in the last but one column.
The rms column marks the mean deviation either from the average value or from the interpolated line.
\end{table}

The range of temperatures and gravities obtained for \iras\, near maximum light
allows us to estimate its spectral type (SpT).
\citet{Kovtyukh2007} adapted the traditional method of the spectral type determination,
that rests on the relative strengths of certain lines, to the echelle spectra of FGK supergiants.
The S/N of our spectrum is not sufficient to apply this method directly,
but we can use a tabulated SpT--$T_{\mathrm{eff}}$ relationship from that study to
estimate the range of SpTs for \iras\, based on its $T_{\mathrm{eff}}$ : F6 (6268 K) -- F9 (5752 K).
In addition, we searched the UVES Paranal Observatory Project (UVES POP)
archive\footnote{http://www.eso.org/sci/observing/tools/uvespop.html} \citep{Bagnulo2003}
for bright field stars with spectral classification adopted from \citet{Buscombe1995}, that would be similar to \iras.
Given the peculiar abundance pattern of \iras\, and
a wide range of line widths exhibited by supergiants,
we did not intend to find a precisely matching standard, but rather wanted to independently
verify SpT of \iras\, implied by its $T_{\mathrm{eff}}$.
The spectrum of HD 108968 (F7Ib/II) appeared to be the closest match to our spectrum of \iras,
hence we adopted SpT F7$\pm$2 I/II for the latter.

\subsection{IR excess}

\begin{figure}
 \includegraphics[width=84mm]{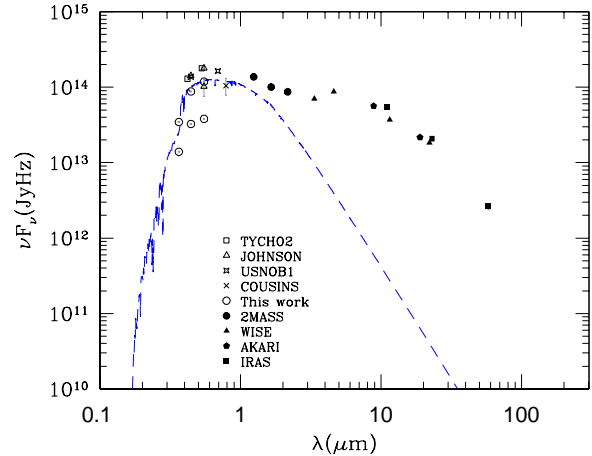}
 \caption{Spectral energy distribution based on our photometry and the measurements from the VizieR database.
 \textit{Dashed line:} Kurucz photospheric model with our spectroscopically determined parameters
 $T_{\mathrm{eff}}$/$\log g$/[M/H]=6000/1.0/$-$1.0,
 reddened by $E(B-V)=0.18$, and shifted to match $V_{max}$ of our epoch of observations.}    
 \label{fig_SED}
\end{figure}

The SED of \iras\, is shown in Fig. \ref{fig_SED}.
Besides our measurements shown for the maximum and minimum light,
we also plot photometry from the VizieR database.
In particular, the measurements in the Johnson's $B$ and $V$ pass-bands originate from
the "All-sky compiled catalogue of 2.5 million stars'' \citep{Kharchenko2009}.
Another measurement in the Johnson's $V$, and a measurement
in the Cousin's $I$ pass-band originate from the ``The Amateur Sky Survey (TASS)
Mark IV patches photometric catalog, version 2'' \citep{Droege2006}.
The rest of the data points arise from the annotated ground and space-based missions.
The scatter in the optical is due to the variability of the source.

Using the photospheric parameters determined in Sec. \ref{chem_comp}),
we can estimate the stellar contribution in the SED and identify any flux excess.
The photospheric contribution in Fig. \ref{fig_SED} is represented by a Kurucz model
\citep{Castelli2003} with $T_{\mathrm{eff}}=6000$ K, $\log g=1.0$, [M/H]$=-$1.0, and $V_{\mathrm{tur}}=2.0$.
The model was reddened by $E(B-V)=0.18$, which was deduced from the comparison of the observed $B-V$ colour in the maximum light
with the expected colour for an F7 supergiant (Sec. \ref{sec_ubv}).
We used reddening law of \citet{Cardelli1989} with $A_{V}/E(B-V)=3.1$
with modification of \citet{Odonnell1994} in the optical to NIR regime.
It is clear that \iras\, has a strong IR excess.
The SED is typical for pAGB stars with dusty discs, where the excess usually starts at 2 $\mu$m,
peaks at around 10 $\mu$m
and drops with a black-body slope towards the far-IR, indicating the presence of large grains \citep{Deruyter2005}.

\section{Photometric variability}\label{sec_ubv}

\begin{figure}
 \includegraphics[width=80mm]{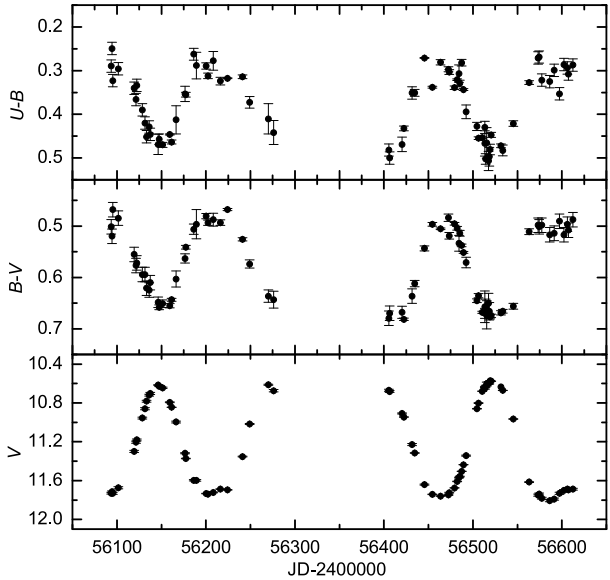}
 \caption{$V$-band light, and $U-B$, $B-V$ colour curves for \iras\,
using our photometry from 2012--2013.}
 \label{fig_UBV}
\end{figure}

\begin{figure}
 \includegraphics[width=83mm]{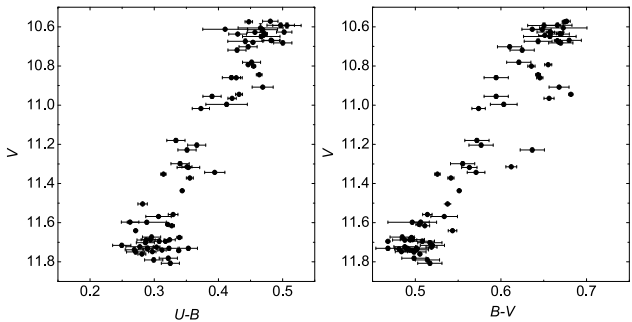}
 \caption{$U-B$ and $B-V$ colour variations with the $V$-band magnitude for \iras\,
in 2012--2013.}
 \label{fig_V_UB}
\end{figure}

In Fig. \ref{fig_UBV} we show our light and colour curves of \iras,
that are clearly variable.
The variations in all three bands are periodic and occur in phase.
Using the \textsc{Period04}\footnote{http://www.univie.ac.at/tops/Period04/}
program of \citet{Lenz2005}, that relies on the discrete Fourier transform for period determination,
we obtained $P=126.53\pm0.40$ d for all three pass-bands.
The peak-to-peak amplitude increases with wavelength:
$\Delta U=1.0$ mag, $\Delta B=1.1$ mag, $\Delta V=1.2$ mag.
The average magnitudes in 2012--2013 were: $<U>=12.17$ mag, $<B>=11.80$ mag, $<V>=11.23$ mag.  
The relationship between brightness and colour is striking (Fig. \ref{fig_V_UB}): 
\textit{when the star fades, it gets bluer}.
Clearly, this behaviour can not be explained simply by variations in the extinction, as it would produce the opposite effect.

To verify how \iras\, behaved in the past,
we searched for time series photometry in the
publicly available catalogues. In 1989--1992 it was observed by \textit{Hipparcos}.
In Tycho-2 \citep{Hog2000} it has the designation TYC 3125-2395-1.
One hundred and eighty measurements are given,
as well as the mean magnitudes $<B_{T}>=10.886$ and $<V_{T}>=10.200$.
Converting to the the Johnson system using the transformation
of \citet{Bessell2000}, we obtain: $<B>=10.75$ mag and $<V>=10.13$ mag.
The errors of the Tycho measurements, however, are too large ($\sigma_{V_{T}}=0.29$ mag and $\sigma_{B_{T}}=0.32$ mag)
for a reliable period determination.

In 2003--2004 \iras\, was monitored by TASS.
Forty-six measurements in Johnson-Cousins $V$ and $I_{C}$ pass-bands
for the first time revealed variability of the star \citep{Sallman2004}.
We obtained the following values of the period for this data set:
127.11$\pm$0.95 d ($V$-band) and 127.11$\pm$0.71 d ($I_{C}$-band).
According to these observations, the $V-I_{C}$ colour did not exhibit sinusoidal variations
with amplitudes larger than the uncertainties of the photometric observations
($\pm$0.07 mag), in contrast to the pronounced $U-B$ and $B-V$ colour variations
observed by us in 2012--2013.

Between May 2006 and January 2008 \iras\, was observed by ASAS.
The measurements are given in ``The catalogue of variable
stars in the Kepler field of view''\footnote{http://www.astro.uni.wroc.pl/ldb/asas/kepler.html}.
The source with a designation 191512+3942.8 has a period $P$=128.8 d,
mean magnitudes $<V>=11.269$ mag, $<I>=10.441$ mag, and the variability type QPER
(a semi-regular variable with a stable period).  
The observations in both bands unfortunately are not simultaneous,
and the behaviour of the $V-I$ colour with phase depends on the details of interpolation.
Nevertheless, the $V-I$ variations did not exceed $\pm$0.04 mag,
in agreement with the TASS observations.
Using the latest data for the smallest aperture size of 30$\arcsec$, which is similar to ours
and avoids contamination from a nearby star, we obtained the period values of 122.06$\pm$0.65 d
for the $V$-band and 127.15$\pm$1.06 d for the $I$-band.
One should note that the quality of the $V$-band measurements in the ASAS catalogue
is very poor -- 91\% of data points have the lowest grade 'D' and only 8\% grade 'A'.
The $I$-band measurements are much better, with 67\% having grade 'A'.
Hence, we adopt $P=127.15$ d for this data set.
  
\begin{figure}
 \includegraphics[width=90mm]{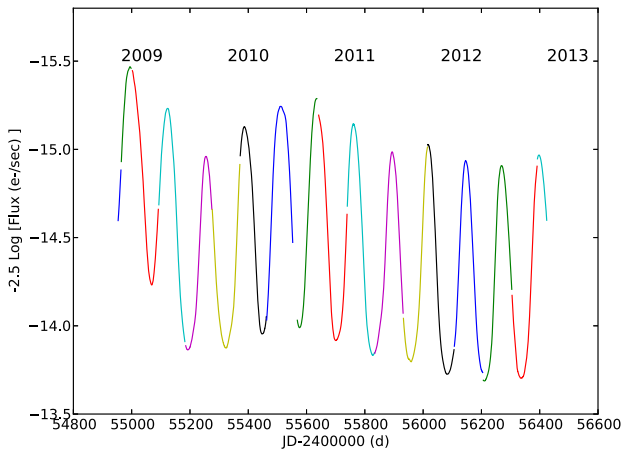}
 \caption{\textit{Kepler} light curve. Colours mark separate acquisition sequences.}
 \label{fig_Kep_JD}
\end{figure}

In Fig. \ref{fig_Kep_JD} we show 4 years of \textit{Kepler} photometry of \iras\, (= Kepler 4644922)
carried out by the \textit{Kepler} satellite.
We plot 18 long-cadence data sets that were available in the STScI archive
at the beginning of 2014.
Due to the fact that the time-scale of variations exceeds the duration of
a \textit{Kepler} observing quarter,
the de-trending procedure performed by the \textit{Kepler} pipeline is invalid.
Raw fluxes are therefore plotted, where small offsets between adjacent
data sets
result from the imperfect calibration.
Applying \textsc{Period04} to these data, we obtain $P=127.497\pm0.014$ d.
The \textit{Kepler} light curve is the most precise of all observations.
It is extremely smooth and shows that the brightness variations are very regular,
but the shape is not fully repeatable from cycle to cycle.

\begin{figure}
 \includegraphics[width=82mm]{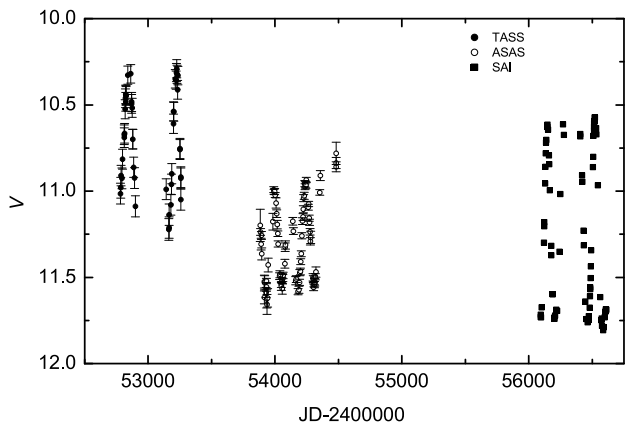}
\caption{Combined $V$-band light curve based on the
TASS, ASAS and our photoelectric photometry (SAI).}
 \label{fig_V_all}
\end{figure}

\begin{table}
\centering
\caption{Summary of the photometric observations}\label{tab_phot_sum}
\smallskip
\begin{minipage}{8cm}
\centering
\begin{tabular}{lccccc}
\hline

survey & $P$ & $<V>$ & $\Delta V$ & $T_{start}$\footnote{First day of observation, JD-2400000} & $\Delta T$ \\
 & (d) & (mag) & (mag) & (d) & (d) \\
 \hline
 TASS    & 127.11$\pm$0.95 & 10.69 & 0.9 & 52782 & 477 \\
 ASAS    & 127.15$\pm$1.06 & 11.36 & 0.7 & 53884 & 598 \\
Kepler   & 127.50$\pm$0.01 & N/A   & N/A & 54953 & 1471 \\ 
SAI      & 126.53$\pm$0.40 & 11.23 & 1.2 & 56093 & 519 \\

\hline
\end{tabular}
\vspace{-0.75\skip\footins}
\renewcommand{\footnoterule}{}
\end{minipage}
\end{table}

The summary of all photometric observations (except Tycho-2) is given in Table \ref{tab_phot_sum}.
As can be seen, there is no indication from this data that the period
changed over the past 10 years.
The shape of the light-curve and the mean magnitude, however, did change from cycle to cycle,
confirming the semi-regular classification of the star. 
In Fig. \ref{fig_V_all} we plot all observations in the $V$-band together.
Tycho-2 measurements have been omitted from the plot due to the large error-bars
and the \textit{Kepler} ones due to a non-standard pass-band.
We performed a period search on this combined data set (200 data points),
by preliminary scaling each of the three data sets to their average magnitude,
and obtained $P=127.04 \pm 2.6$ d, which, as will be shown in the next section,
coincides with the spectroscopic period.

\begin{figure}
 \includegraphics[width=75mm]{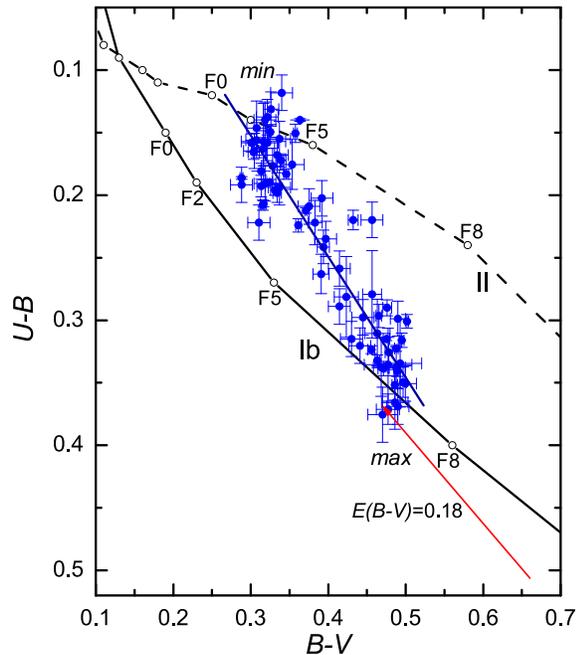}
 \caption{Location of \iras\, on the $U-B$ vs. $B-V$ diagram,
de-reddened by $E(B-V)=0.18$ and fitted with a straight line to guide the eye.
The corresponding reddening vector is shown as a red line,
following equation $E(U-B)/E(B-V)=0.72+0.05E(B-V)$.
Black solid and dashed lines designate sequences of supergiants (Ib) and bright giants (II),
respectively.}
 \label{fig_UB_BV}
\end{figure}

Our simultaneous $UBV$ photometry enabled us to study
colour variations in \iras\, for the first time.  
Fig.~\ref{fig_UB_BV} shows an area on the  $U-B, B-V$ two-colour diagram traversed by \iras.
The sequences of supergiants and bright giants according to \citet{Schmidt-Kaler82} are drawn for reference.
From our spectroscopic analysis it follows that at maximum light
the star should be of spectral type F7 ($\pm2$). According to the
supergiant calibrations of \citet{Schmidt-Kaler82} and \citet{Kovtyukh2007},
this corresponds to $(B-V)_{0}=0.48\pm0.1$ mag. The observed $(B-V)$ at maximum light
is $\sim$0.66, hence $E(B-V)=0.18\pm0.1$ mag.
Assuming that this reddening is predominantly interstellar,
and adopting the standard reddening law of \citet{Hiltner1956}: $E(U-B)/E(B-V)=0.72+0.05E(B-V)$,--
we have de-reddened all points in Fig.~\ref{fig_UB_BV} by this value.
As can be seen, the colours vary considerably throughout the cycle,
but follow neither reddening nor temperature sequences. 
While in the maximum light both colours can be explained by a moderately reddened F7 Ib supergiant,
in the minimum light the star can not be brought to the same locus
on the UBV diagram even when allowing for a different amount of reddening.
Pulsations can hardly explain this behaviour either, as such stars normally get redder at minimum light.
Could the blue colours at minimum light be due to the contribution of
an early-type companion?
In the next section we will examine our time series of spectra
to investigate this possibility.

\section{Spectroscopic variability}

\subsection{Radial velocities}\label{sec_rv}

To determine RV of \iras, we used a cross-correlation method with
a mask containing lines of a G2 star.
The correlation was performed on the order-by-order basis, after which
the cross-correlation functions (CCFs) from the central 20 orders were merged
into one to represent an average line profile.
Normally, this profile would be fit with a Gaussian to obtain an RV measurement.
However, in many phases the CCFs turned out to be of a more complex shape.
While some CCFs have the form of a single absorption component, either strong and narrow
or shallow and broad, in other CCFs two separate components can be clearly seen.
The determination of the CCF centroid is therefore not straightforward.

\begin{figure}
 \includegraphics[width=84mm]{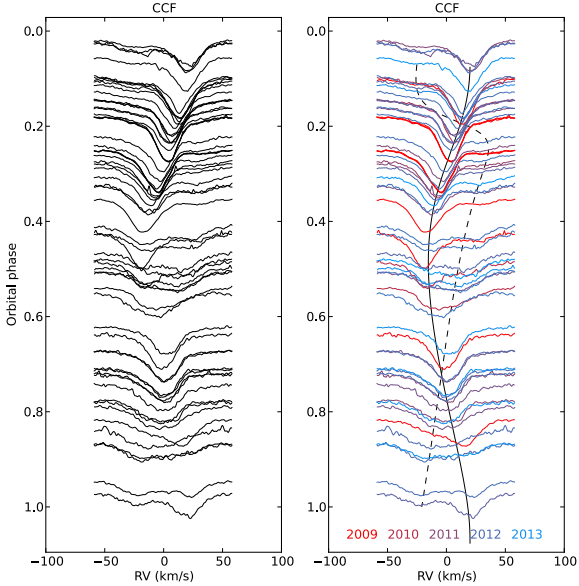}
 \caption{Cross-correlation functions arranged according to the orbital phase, with time running downwards.
Throughout the paper the phases have been defined as follows: $\phi=(JD-2454980.42464)/127.08$,
with $\phi=0$ corresponding to $RV_{max}$.
The colours on the right panel designate the year of observation
to illustrate the cycle-to-cycle variations.
The solid and dashed lines trace the main
and the secondary CCF components, respectively.}
 \label{fig_ccf_prof}
 \end{figure}

To investigate whether this behaviour could be periodic,
we first measured a flux-averaged value of the RV for
each CCF, using $(1-F_{\lambda})^{2}$ as a weighting function.
Applying \textsc{Period04}
to these data revealed several peaks on the periodogram,
of which the first two strongest ones are at 63 d and 129 days.
The latter coincides with the photometric period,
while the former reflects the fact that the lines split twice per photometric period.
Arranging the CCFs according to the phase calculated with the longer period
allowed us to identify the main component (which is usually the stronger one) and the secondary component
in the split CCFs (Fig. \ref{fig_ccf_prof}).
We then re-determined RVs separately for each component using a double-Gaussian fit.
In several cases it was not possible to identify the secondary component,
because it was in the noise or merged with the main component.
In such cases a single-Gaussian fit was performed, and the RV was assigned to the main component.
The resulting RVs are given in Table A2.

\begin{figure}
 \includegraphics[width=84mm]{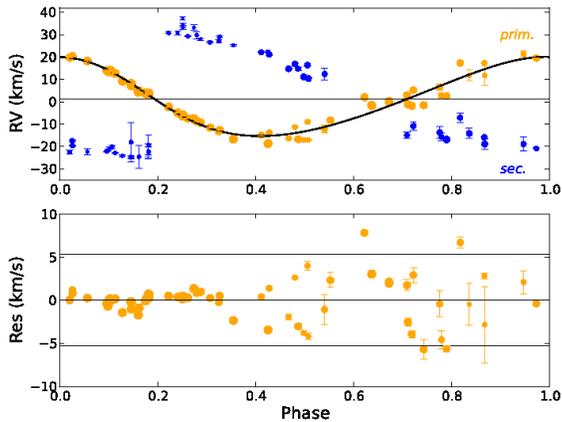}
 \caption{Radial velocities for the main component of the CCF (orange circles) and for the secondary component (blue).
The former were fit with a Keplerian orbit, which is drawn in the upper panel,
while the residuals from it are plotted in the lower panel. 
The size of the circles is proportional to the component's strength.
Horizontal lines mark $\pm 3 \sigma$ deviation from the fit orbit.}
 \label{fig_ccf_fit}
 \end{figure}

\begin{figure}
 \includegraphics[width=84mm]{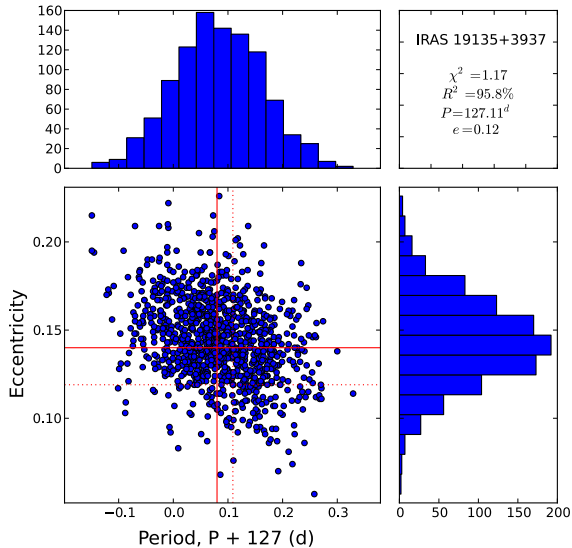}
 \caption{Distribution of values for eccentricity and orbital period
 based on 1000 Monte-Carlo simulations of the RVs of the main CCF component.
 Solid lines mark a Keplerian solution obtained from the observed RVs,
 whereas the dotted lines denote a simulated RV set that resulted in the smallest $\chi^{2}$ value. }
 \label{fig_orb_pars}
 \end{figure}

In order to determine whether the main RV component could be due to the orbital
motion of the brighter star (the primary) in a binary system,
we fit it with a Keplerian orbit, which included a re-evaluation of the period.
The fit is very good -- see Table \ref{tab_orbit_pars} and Fig. \ref{fig_ccf_fit}.
The larger residuals from the fit between phases 0.4 -- 0.9
are due to the broader CCFs.
To estimate the uncertainties on the orbital parameters,
we carried out 1000 Monte-Carlo realisations to simulate the observed RVs.
For each date we used a Gaussian distribution centred on the observed RV
and $\sigma$ adopted as follows: 0.7 km~s$^{\mathrm{-1}}$ for the orbital phases between $\phi=0.95-1.35$
and 3.5 km~s$^{\mathrm{-1}}$ for $\phi=0.35-0.95$, based on the residuals from the Keplerian fit.
For each realization the orbital parameters were determined,
and the $\sigma$ of the distribution of a given parameter was adopted as the parameter's uncertainty.    
Fig. \ref{fig_orb_pars} illustrates this procedure for the period and eccentricity.

The RV curve of the secondary CCF component is much less certain,
in particularly between phases 0 $-$ 0.3, so we did not attempt to model it.
However, we can use the deduced mass function from the fit to the RV of the main component
to estimate the mass of the putative companion.
Adopting 0.5 M$_{\odot}$ for the mass of the pAGB primary (a mass of a typical white dwarf),
with the mass function of 0.07 M$_{\odot}$
we obtain masses of the companion between 1.1 and 0.4 M$_{\odot}$, depending on the adopted inclination
of $30\degr$ and $>65\degr$, respectively. The companion has a comparable or larger
mass than the primary.

\begin{table}
\centering
\caption{Orbital elements for the primary (pAGB) component of \iras.}             
\label{tab_orbit_pars}
\begin{tabular}{lcc}
 \hline     
Parameter  & Value  &  $\sigma$  \\
\hline 
$P$ (d)            &  127.08    & 0.08 \\
$a$ $sini$ (AU)    &  0.20     & 0.007 \\
$f(m)$ ($\mathrm{M_{\odot}}$) &  0.07       & 0.007 \\
$K$ (km~s$^{\mathrm{-1}}$)         & 17.71      & 0.5 \\
$e$                & 0.14       & 0.02 \\
$\omega$ ($\degr$) &  66       & 14 \\ 
$T_{\mathrm 0}$ (JD)        & 2 454 997.8 & 6.8 \\
$\gamma$ (km~s$^{\mathrm{-1}}$)    & 1.3    & 0.4 \\
$\chi_{red}^{2}$  & 1.11 & \\
$R^{2}$ & 95.4\% & \\
 \hline                  
\end{tabular}

\medskip
Listed are orbital parameters with their uncertainties, 
reduced chi-square, and the coefficient of determination. 
\end{table}

\iras\, turned out to be yet another disc system that is a binary. 
The line behaviour is reminiscent of the double-lined
spectroscopic binaries, but not identical: the relative strengths of the
two CCF components in our case vary, both with phase and from cycle to cycle.
While the main component can be safely attributed to the pAGB star,
the association of the secondary component with a physical companion is not straightforward.

\begin{figure}
 \includegraphics[width=90mm]{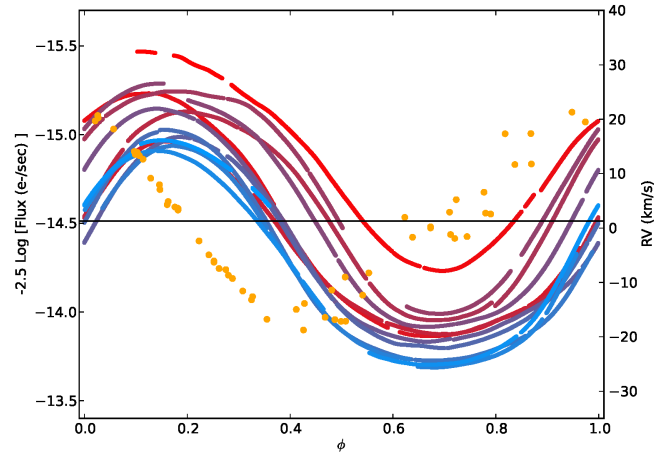}
 \caption{Radial velocities (orange circles) and \textit{Kepler} fluxes (red to blue points) plotted against the orbital phase.
The colour scheme is the same as in Figs. \ref{fig_ccf_prof} and \ref{fig_Hal_1d}. The horizontal line marks
the systemic velocity.}
 \label{fig_RV_Kep}
 \end{figure}

In Fig. \ref{fig_RV_Kep} we overlay RVs of the main CCF component and the \textit{Kepler} photometry,
that were obtained over the same time-span.
Despite the cycle-to-cycle variations in the light curve,
an offset of a quarter of a period can be clearly seen with the RV curve:
the superior conjunction of the primary star ($\phi \sim 0.2$) coincides with the maximum light,
while the inferior conjunction ($\phi \sim 0.7$) with the minimum light.
Brightness declines are therefore consistent with the obscuration of the primary
by the inner disc wall.

Over half of the orbit centred on the minimum light ($\phi=0.4-1.0$)
one also observes an increase in the RV scatter.
As can be seen in Fig. \ref{fig_ccf_prof}, this is not due to the decreased S/N, but due to the fact
that the secondary component of the CCF becomes of comparable strength to the main component,
resulting in the shallower, smeared combined profiles that are difficult to fit.
This is not surprising considering that at $\phi=0.7$ the primary becomes obscured,
and the spectrum of the companion should become more visible.
In Sec. \ref{sec_refl}, however, we will present some arguments
that hinder a definite identification of the secondary CCF component
with the spectrum of the companion.

\subsection{Emission lines}\label{sec_eml}
\begin{figure*}
   \centering
     \includegraphics[width=8.4cm]{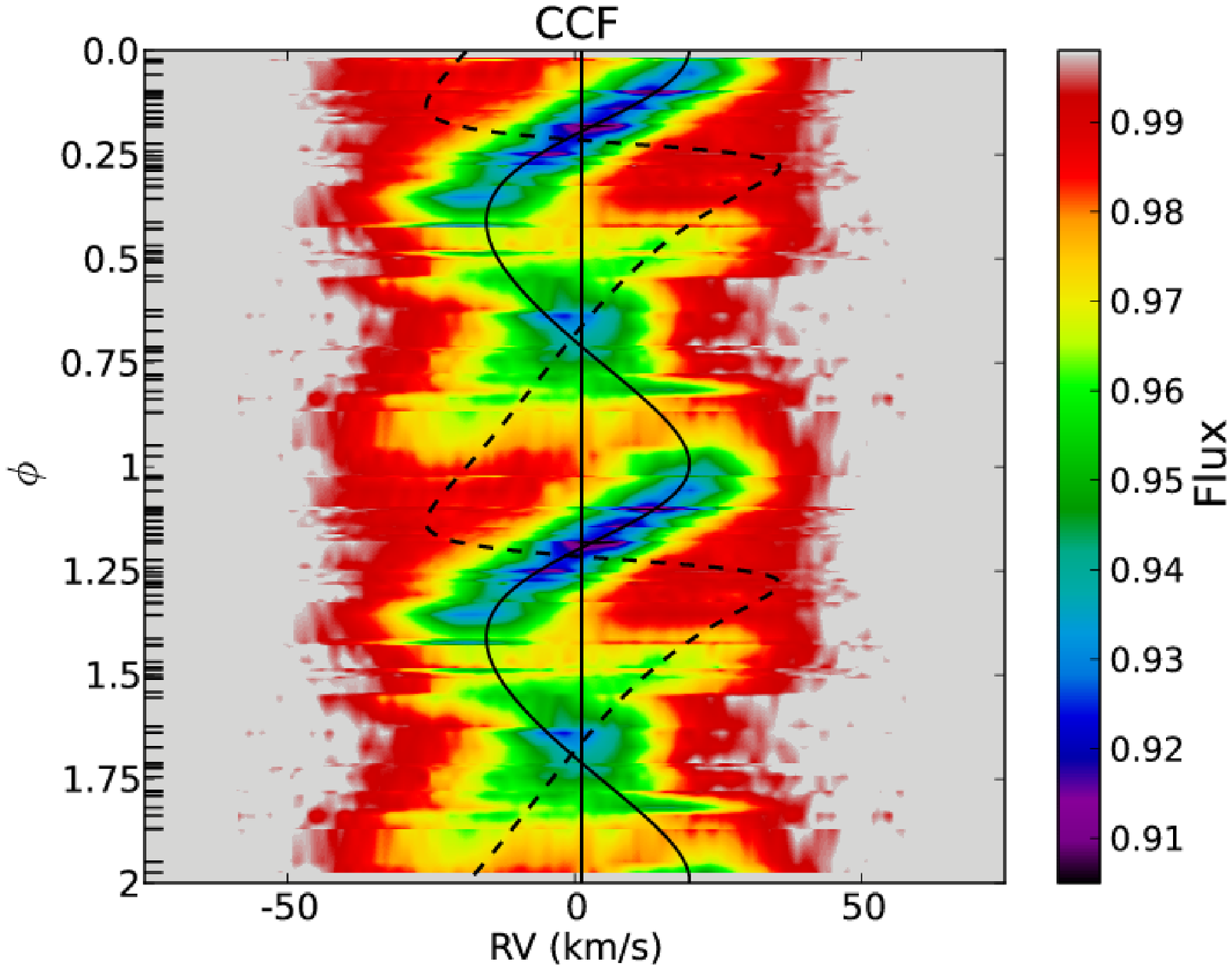}
     \includegraphics[width=8.4cm]{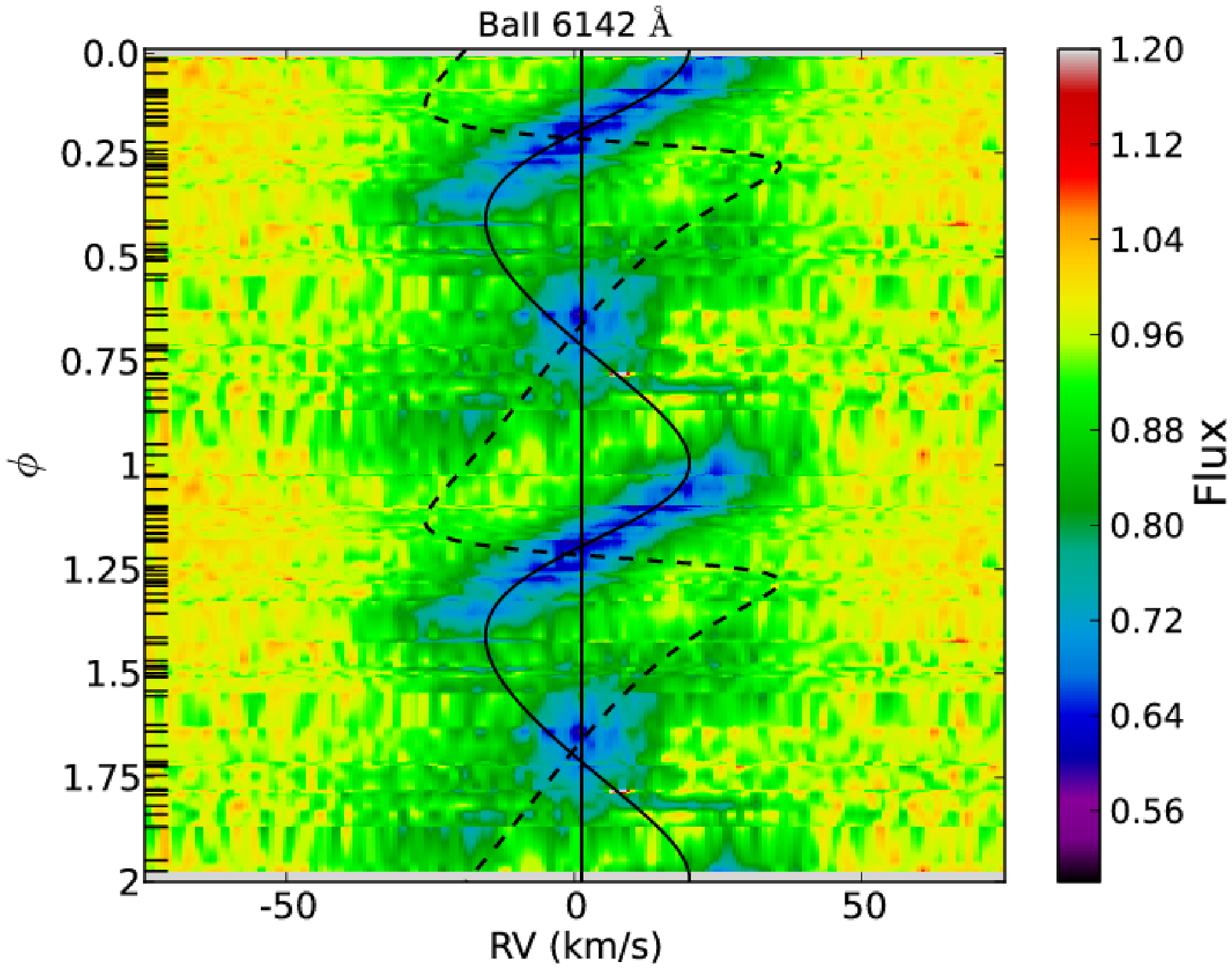}
     \includegraphics[width=8.4cm]{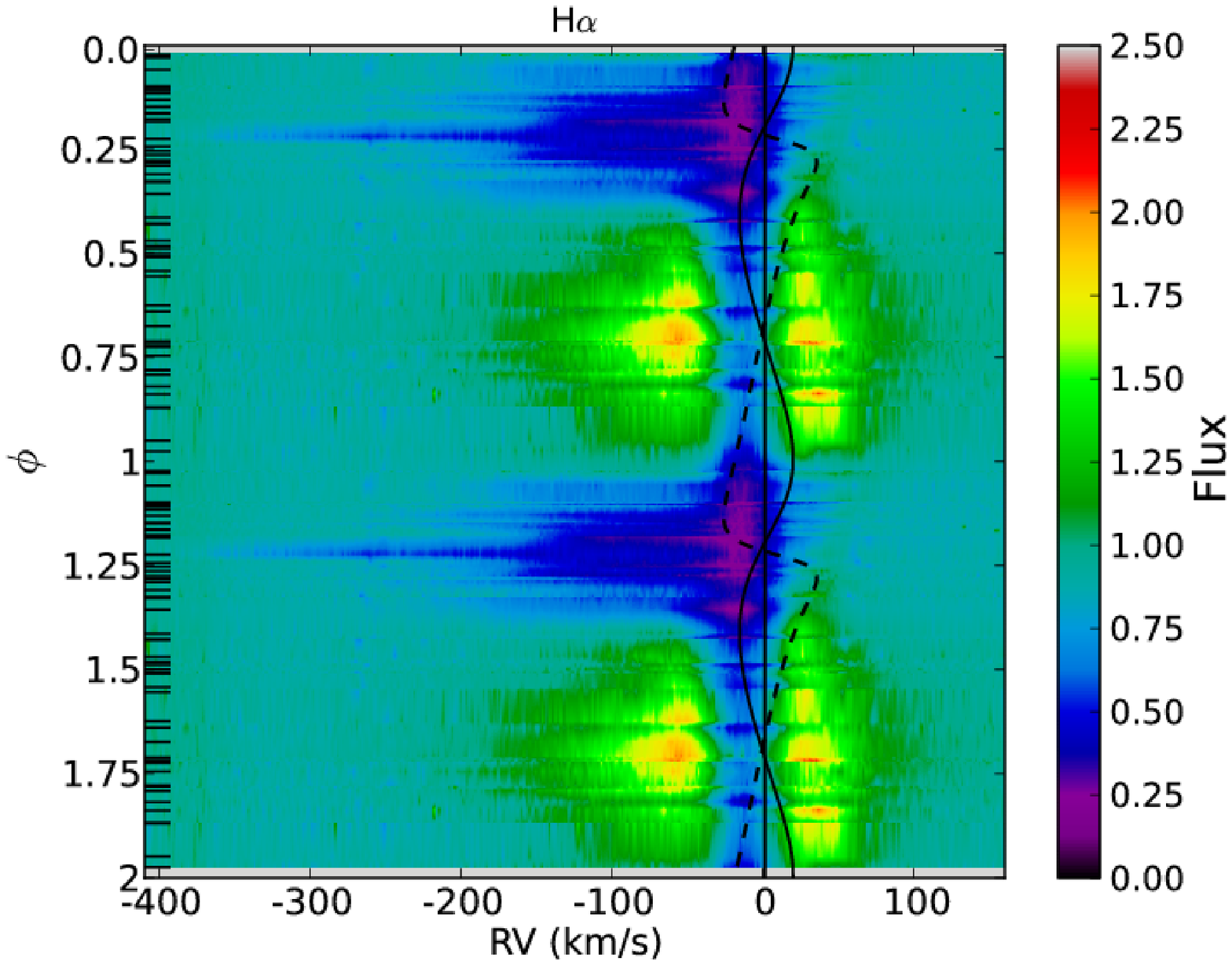}
     \includegraphics[width=8.4cm]{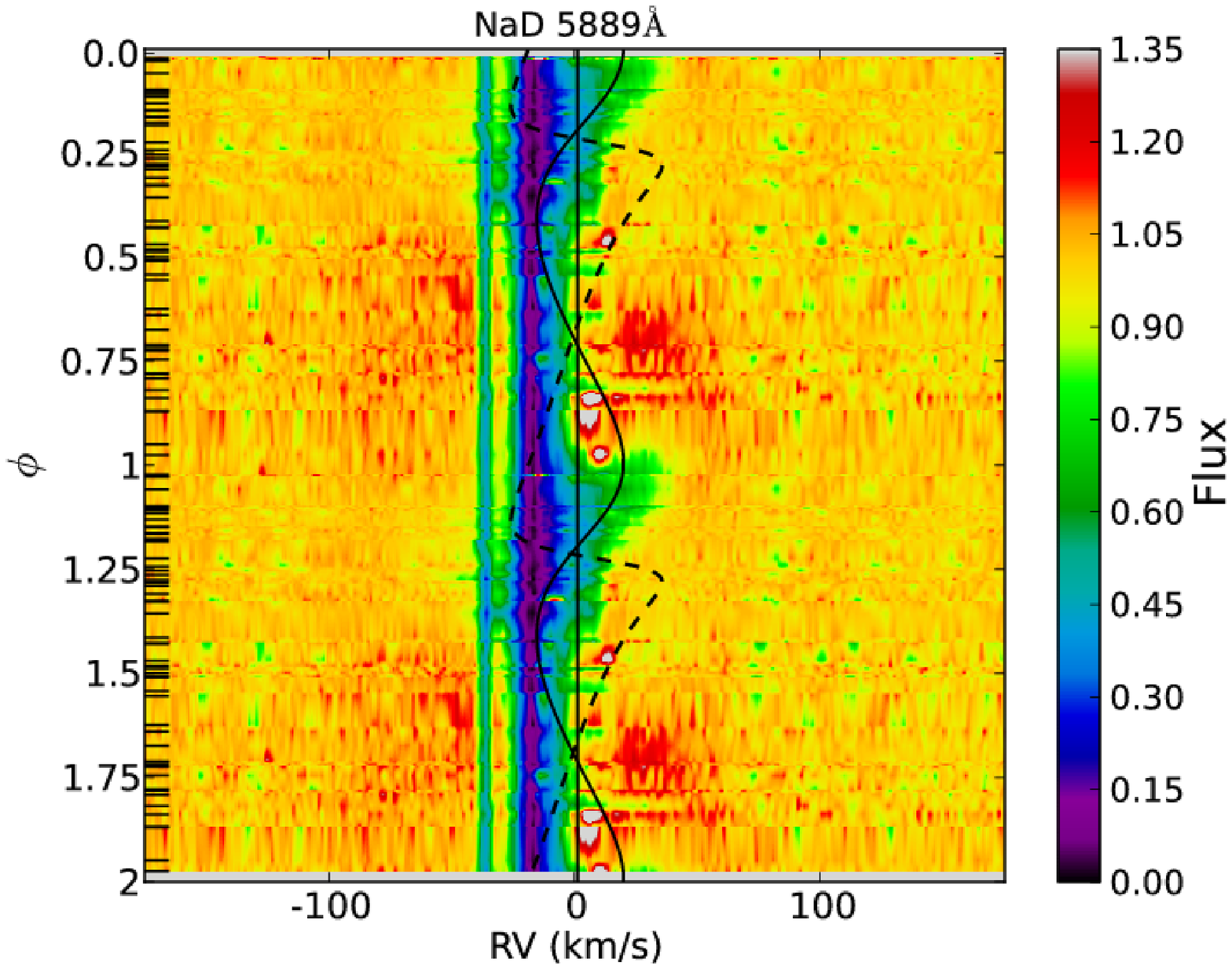}
     \includegraphics[width=8.4cm]{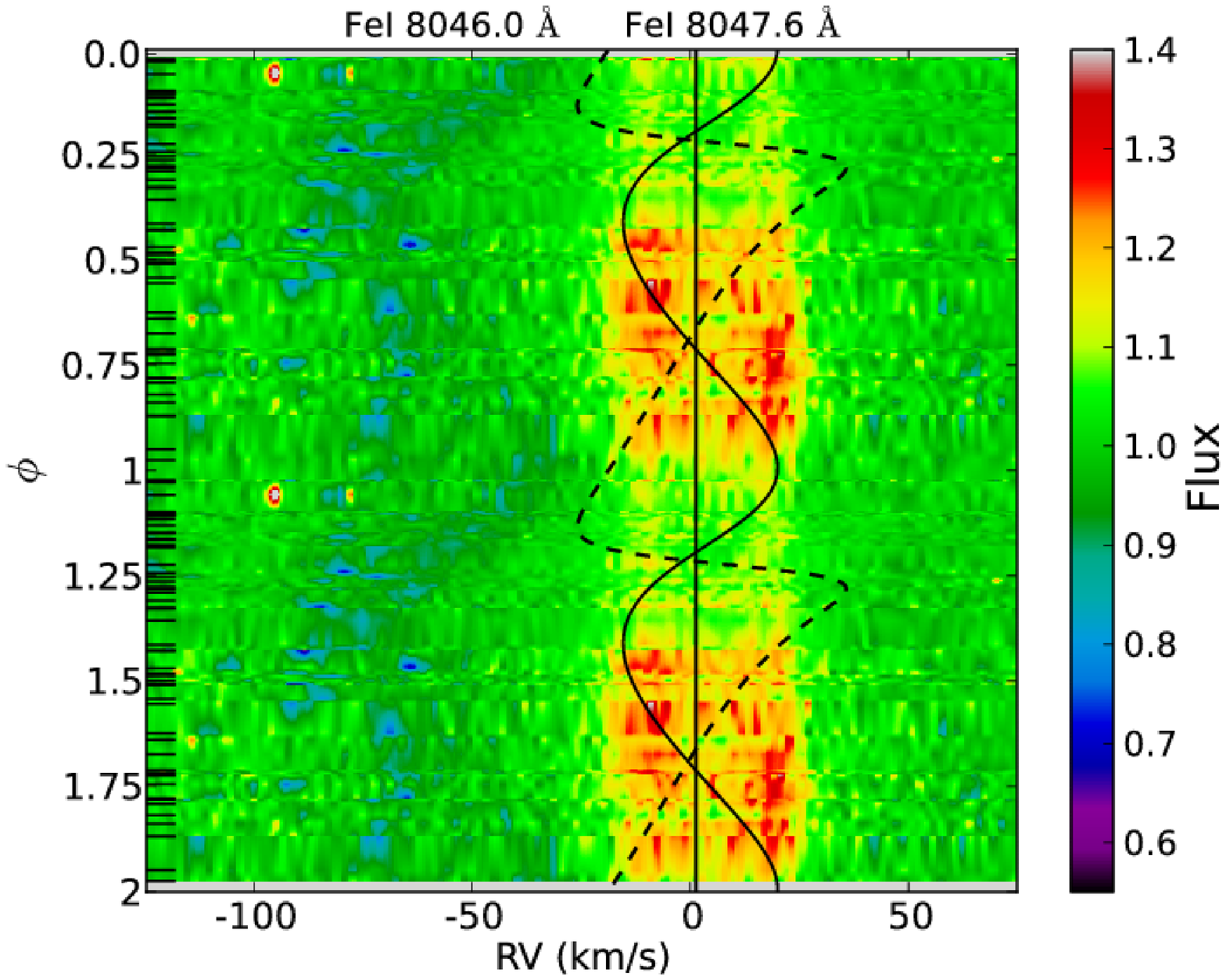}
     \includegraphics[width=8.4cm]{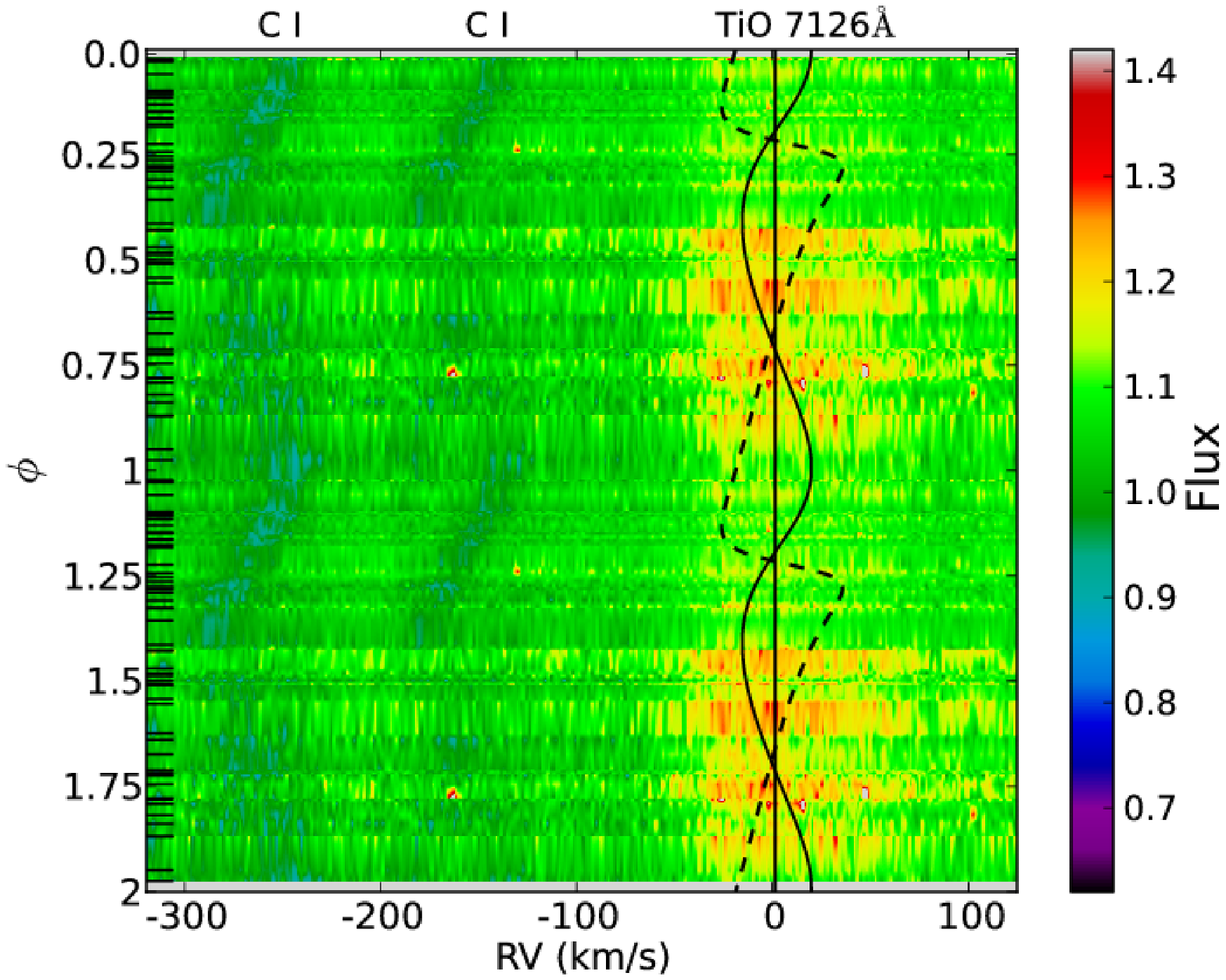}
 \caption{Dynamic spectra of the representative features in the spectra of \iras:
top panel shows the behaviour of the photospheric absorption lines, middle panel - lines dominated
by the circumstellar emission and absorption, bottom panel - pure emission lines (with the neighbouring
photospheric lines of FeI and CI shown for reference).
Colours represent continuum-normalized fluxes, with unity corresponding to the continuum level.
Spectra are arranged according to the orbital phase with period 127.08 d; one period is shown twice; time runs down.
The following lines are drawn to guide the eye: vertical line for the centre-of-mass velocity, 
solid curve -- for the Kepler's solution of the main CCF component, dashed curve -- of the secondary component.
Short horizontal dashes designate observed phases.}
 \label{fig_2d}
 \end{figure*}

There are a number of variable emission lines in the spectrum of \iras.
Fig. \ref{fig_2d} presents an overview of the discussed features in the form of dynamic spectra.
The upper panel shows the behaviour of the photospheric lines
using the example of the cross-correlation function and one strong Ba\,\textsc{ii} line;
the middle panel shows emission-absorption profiles of H$\alpha$ and one component of the NaD doublet;
the bottom panel compares static emission lines of Fe and TiO with
the nearby dynamic photospheric lines.

\begin{figure}
 \includegraphics[width=84mm]{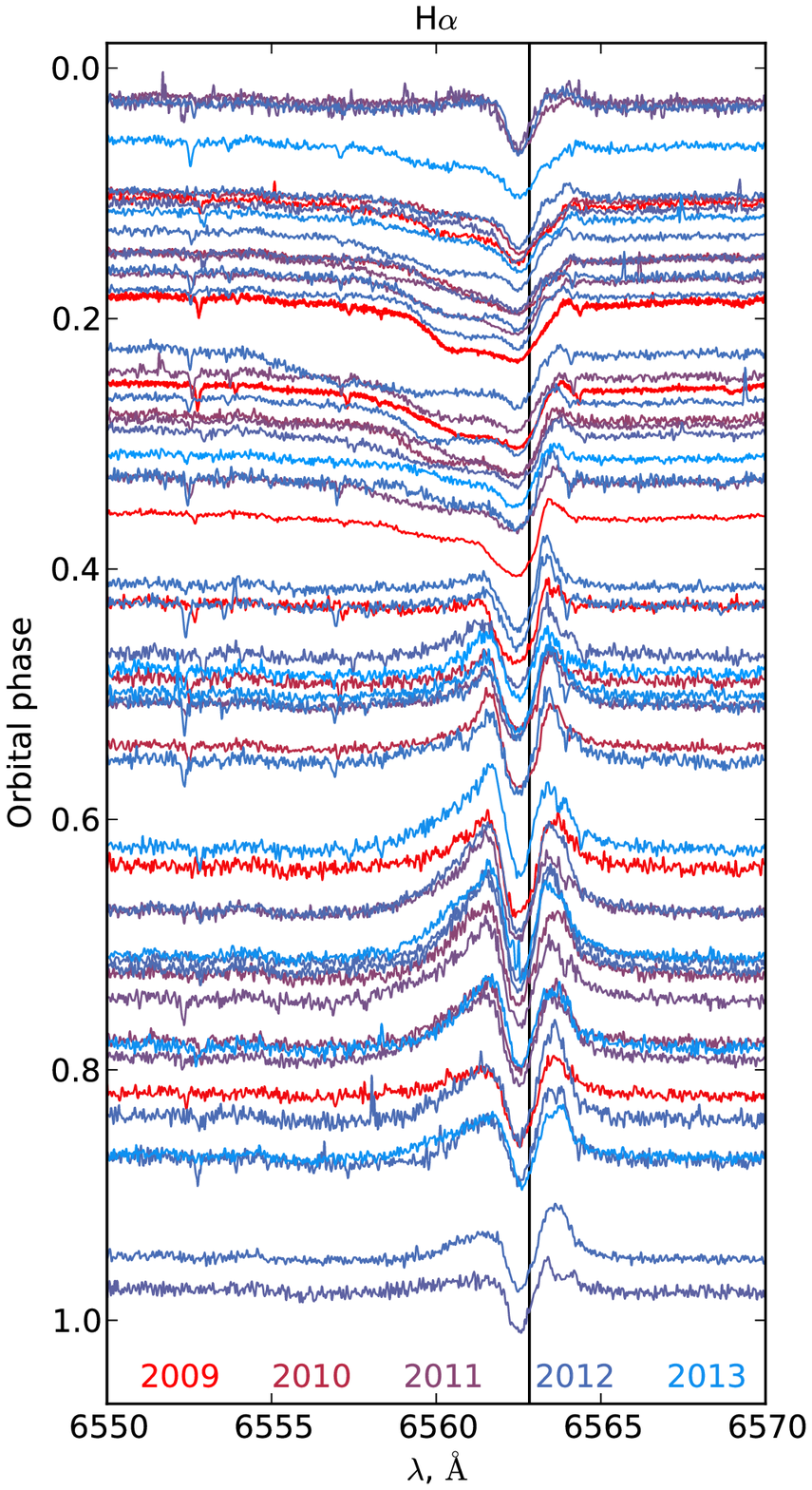}
 \caption{H$\alpha$ as a function of the orbital phase.
 Vertical line marks the systemic velocity.}
 \label{fig_Hal_1d}
 \end{figure}

The most prominent emission line is H$\alpha$
(higher members of the Balmer series show a similar behaviour, but to a much smaller degree
and confined to the line core).
H$\alpha$ profiles, arranged according to the orbital phase, are shown in Fig. \ref{fig_Hal_1d}.
Over one half of the period, centred on $\phi=0.7$ (pAGB inferior conjunction), the profiles take a shape of
a double-peak emission, with a distance between the peaks of $\sim$100 km~s$^{\mathrm{-1}}$
and a width at the base of at least 300 km~s$^{\mathrm{-1}}$ (before photospheric absorption subtraction).
Alternatively, such profiles can be interpreted by a broad emission
profile with a narrower superimposed absorption component.
This absorption can not be photospheric due to the fact that it does not follow
the orbital motion, but is permanently 13 km~s$^{\mathrm{-1}}$ blue-shifted relative to the systemic velocity.
Over another half of the period the emission in H$\alpha$
is overtaken by a broader, asymmetric absorption,
with a blue wing that extends to over 350 km~s$^{\mathrm{-1}}$ near $\phi=0.2$ (pAGB superior conjunction). This behaviour is identical to that in \bd\, \citep{Gorlova2012a}.

\begin{figure*}
   \centering
 \includegraphics[width=55mm]{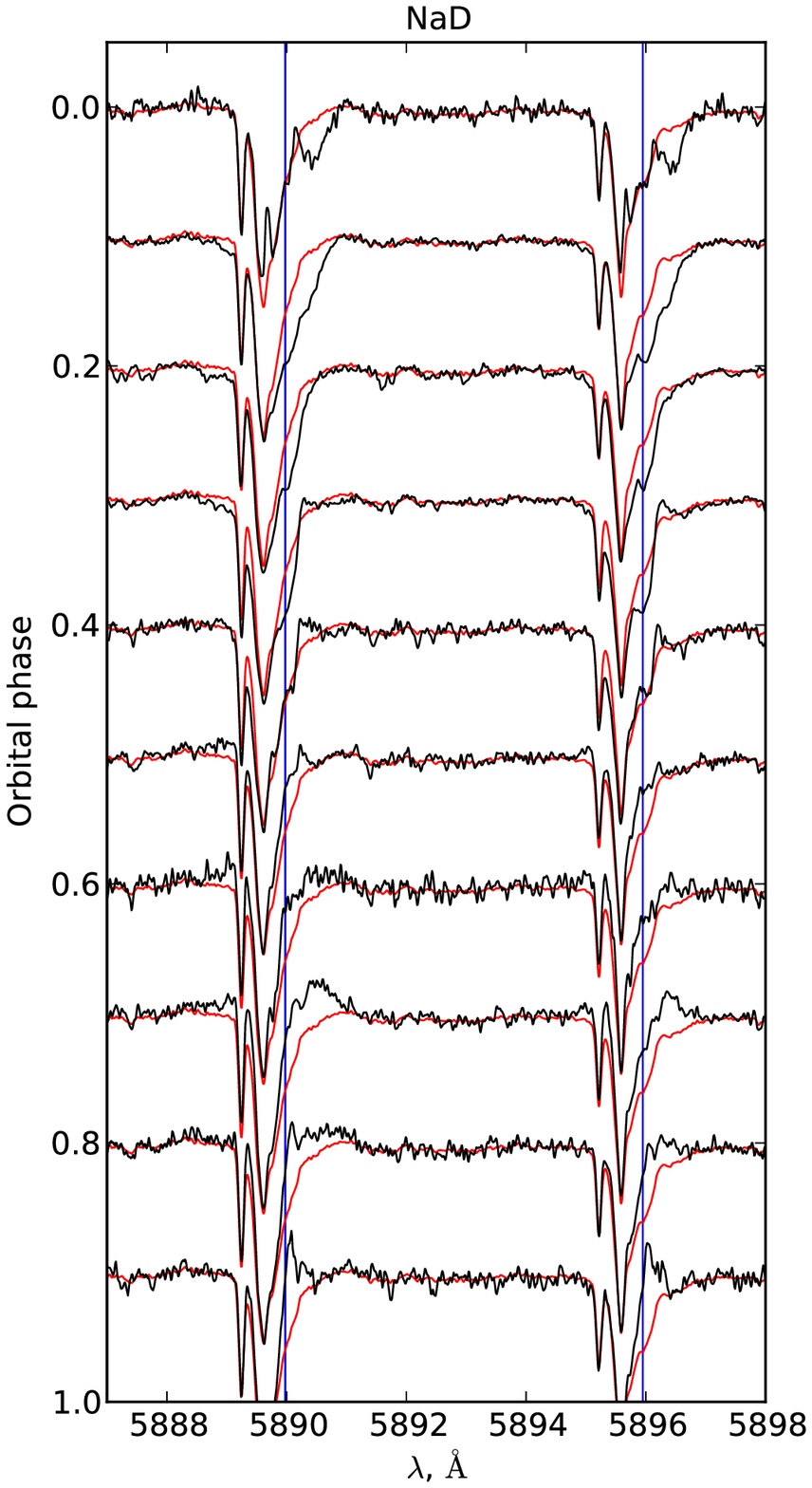}
 \includegraphics[width=55mm]{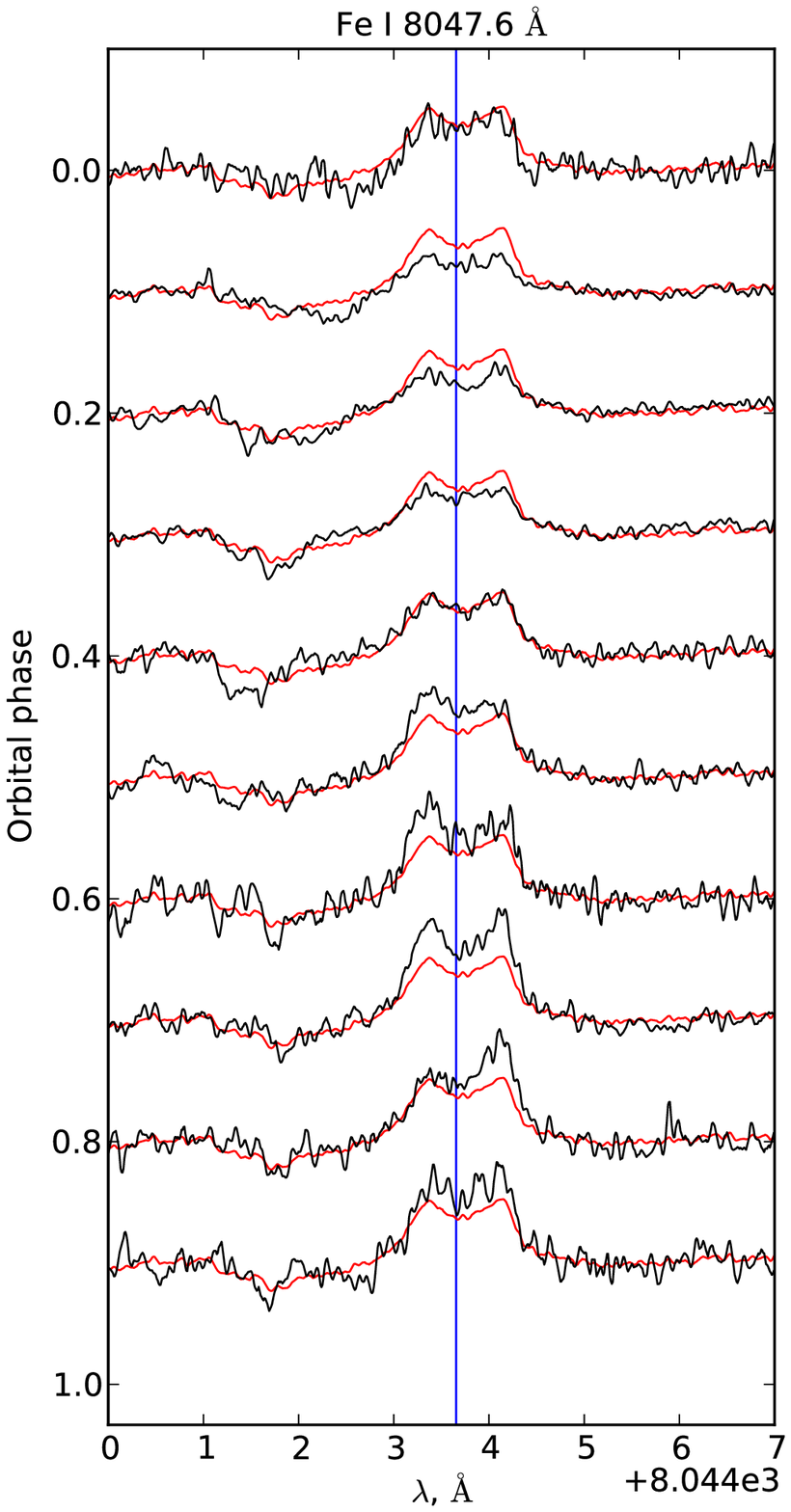}
 \includegraphics[width=55mm]{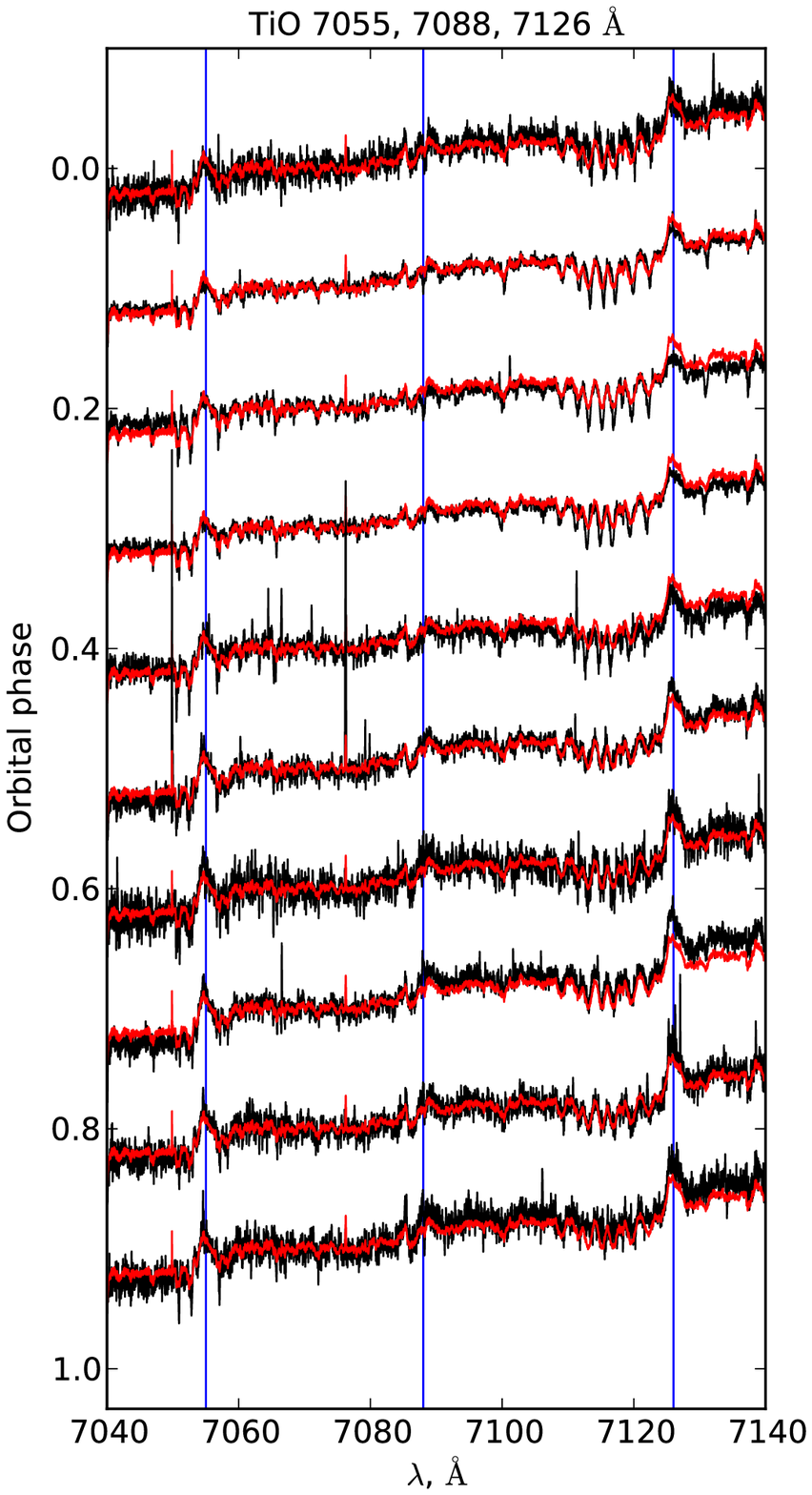}

 \caption{Emission lines as a function of orbital phase. {\it Black:} an average of all spectra falling within
a given phase bin; {\it red:} an average spectrum over the entire orbit.
All spectra have been normalized to unity at the continuum and shifted vertically according to the orbital phase.
Vertical lines mark wavelengths for the identified features at the systemic velocity.}
 \label{fig_sp_av}
 \end{figure*}

In some spectra we also spotted weak emission in a number of low-excitation metallic and molecular lines.
To study a possible correlation with the orbital phase,
we first averaged spectra within each of 10 phase bins to increase the S/N.
These combined spectra arranged according to the phase are shown in Fig. \ref{fig_sp_av}.
Three examples of emission features are given: the resonance lines of Na\,\textsc{i},
low-excitation Fe\,\textsc{i} line 8047.62 \AA\, ($\chi_{low}-\chi_{up}=0.9-2.4$ eV),
and three close TiO band-heads. 

The Na D lines in \iras\, are dominated by a composite ``shell'' absorption component,
which is common for pAGB stars.
The photospheric component is only visible around $\phi=0$, when it is maximally redshifted.
Throughout the rest of the orbit it is hidden in a cluster of narrow absorption lines,
that are permanently blue-shifted relative to the systemic velocity.
Different features can be better traced in the dynamic spectra, as  
shown in Fig. \ref{fig_2d} (middle right panel). There, the Na D photospheric absorption can be traced
up to $\phi=0.4$, which corresponds to the least obscured part of the primary's orbit,
when other photospheric lines are strongest as well.
At least two non-photospheric absorption components can be disentangled: at $-38$ and $-20$ km~s$^{\mathrm{-1}}$,
that can be both of inter- and circumstellar origin.
What is not so common is the appearance of the broad emission wings between $\phi=0.6-0.8$
(in Fig.\ref{fig_2d} they can be seen as a flux excess over the mean profile),
that are reminiscent of emission in H$\alpha$. 

Some weak emission lines are found in the red part of the spectrum,
where they better stand out against the dropping continuum.
The three most notable lines can be associated
with two close low-excitation multiplets ($\chi_{up}=2-3$ eV) of Fe\,\textsc{i},
with rest wavelengths of 6400.32, 6498.94, and 8047.62 \AA\,.
Many more can be hidden in the telluric features and noise.
Careful examination reveals that they have double-peak,
but relatively narrow, profiles (the distance between the peaks is
$\sim$ 30 km~s$^{\mathrm{-1}}$),
that are permanently centred on the systemic velocity (within 3 km~s$^{\mathrm{-1}}$).
The latter two lines were also noted by us in \bd, but in the
case of \iras\, they are stronger and clearly variable.
In the continuum normalized spectra,
they appear to vary in strength with the photometric/orbital period.
They are 2.5 times stronger in the minimum light than in the maximum,
which coincides with a one magnitude amplitude of the photometric variations.
This means that the line emission flux is constant,
and the apparent line variability is due to the variations in the continuum flux.

Finally, we detected weak emission from three TiO band-heads belonging to the $\gamma(0,0)$
system. Similarly to the atomic emission lines,
they become strongest during the minimum light, and appear to be stable in the RV.
The individual transitions, however, are blended with each other,
so it is not possible to gauge the kinematics of the emitting region.

\section{Discussion}

We showed that \iras\, possesses typical characteristics of
a pAGB star with a dusty disc: it is a low-gravity F star
with large IR excess but low reddening, it is deficient in refractory elements,
and it has a strong and variable H$\alpha$ profile indicative of mass loss.
In particular, it is very similar to \bd\, described in our earlier work
\citep{Gorlova2012a}. The only substantial difference between the two is that \iras\,
is a recognized variable star due to the larger amplitude of the light curve.
The HERMES survey uncovered periodic, but complex spectral behaviour in both stars.
In the following subsections we will elaborate upon our model
of an interacting binary system that we developed
for \bd\, to explain additional aspects exibited by \iras:
brightness variations (Sec. \ref{subsec_var}), stationary emission lines (Sec. \ref{subsec_accr}),
photospheric line splitting and peculiar colour variations (Sec. \ref{sec_refl}).

\subsection{Binarity and SRd variability}\label{subsec_var}

For the first time we have demonstrated that \iras\, is also variable in radial velocity,
with a period coinciding with the photometric period.
We can now use the same argument as for \bd\, to
disregard pulsations as the cause of variability.
Integrating the RV curve of the main CCF component over half of the
period, we obtain a displacement $\sim$90 R$_{\odot}$,
which is comparable with the radius of the pAGB star itself and would be prohibitively large
for pulsations. And if line splitting were due to the propagation of shock waves,
as in RV Tau stars, one would have to explain why this star 
does not show other characteristic features of these variables,
such as interchanging deep and shallow minima,
and, most notably, variations in the effective temperature.

\begin{figure}
 \includegraphics[width=84mm]{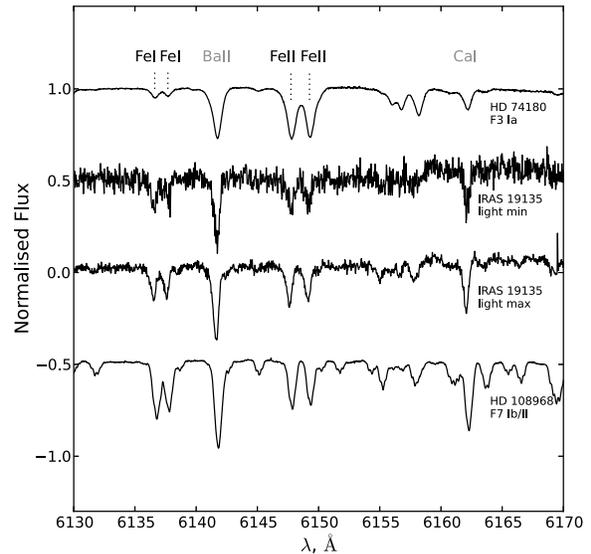}
 \caption{Comparison of \iras\, spectra in the phases of minimum and maximum light
 with each other and with two spectral standards.}
 \label{fig_sp_comp}
 \end{figure}

The constancy of $T_{\mathrm{eff}}$ is demonstrated in Fig. \ref{fig_sp_comp},
where we compare a part of \iras\, spectrum in the phases of minimum and maximum light
with the spectra of two standard F supergiants from the UVES POP archive: HD 74180 (F3 Ia, \citealt{Malaroda1975})
and HD 108968 (F7 Ib/II, \citealt{Houk1975}).
The standards were chosen to have SpTs expected for
\iras\, based on its $B-V$ colour in those phases.
It can be seen that in the late-F standard
the lines of neutral species become noticeably stronger 
than in the early-F one.
In the spectra of \iras, however, this effect is not observed,
hence, the temperature must have remained constant between the two opposite phases.
Another confirmation of this fact is that \citet{Rao2014}
studied \iras\, in the intermediate phase ($\phi=0.49$)
and obtained precisely the same value of $T_{\mathrm{eff}}$ as us.

We conclude that the RV variations in \iras\, are due to the orbital motion
of the giant primary around a much fainter companion,
which is very common among pAGB stars with discs.
The periodic semi-regular dimmings can be explained
by the obscuration of the primary by the circumbinary disc matter,
most likely by the puffed-up inner rim during the inferior conjunction.
The absence of a flat part on the light curve near maximum light
could be due to the permanent obscuration or
presence of a large amount of scattered light.
This explanation has been already proposed for a few other long-period pAGB variables
with the same phase shift between the light and RV curves,
such as HR 4049 \citep{Waelkens1991},
HD 52961 \citep{VanWinckel1999}, and EN TrA \citep{VanWinckel2009}.
The very smooth Kepler light curve indicates that the
photosphere of \iras\, is very stable.

\subsection{Accretion onto companion and gas discs}\label{subsec_accr}

Furthermore, a number of spectroscopic features in \iras\, point
to the presence of an \textit{active mass transfer} between the primary and companion.
In particular, as was described in Sec. \ref{sec_eml},
near $\phi=0.25$ H$\alpha$ develops a spectacular P Cyg-like profile.
With our phase convention, where phases are counted from the time of the maximum RV of the primary,
$\phi=0.25$ for a nearly-circular orbit corresponds to the giant's superior conjunction
(where it is farthest away from us, while the putative companion is in-between).
In \citet{Gorlova2012a} and \citet{Gorlova2013} we describe a few other systems
from the HERMES survey and from the literature with a similar behaviour of H$\alpha$.
As was shown by \citet{Thomas2013} for the central star of the RR nebula,
this phenomenon can be explained by a wide-angle jet originating
at the secondary. The lobe pointing in our direction will be periodically projected
against the giant primary and produce the observed transient blue-shifted absorption.
The jet is likely powered by accretion from the pAGB primary to the secondary.

The interpretation of the double-peaked emission lines is not so straightforward.
Using STIS spectrograph on the \textit{Hubble Space Telescope},
\citet{Thomas2011} spatially resolved narrow 
emission (with the distance between the peaks of 12
km~s$^{\mathrm{-1}}$) in the NaD lines of the RR,
and deduced that it is produced in the distant parts of a bipolar outflow,
that are seen in the direct light.
On the other hand, the much broader emission in H$\alpha$
according to \citet{Witt2009} could form in the parts of the lobes
that are closer to the binary.
The RR, however, is not a typical disc object, and the appearance of some features
may be affected by its nearly perfect edge-on orientation.

\begin{figure}
 \includegraphics[width=84mm]{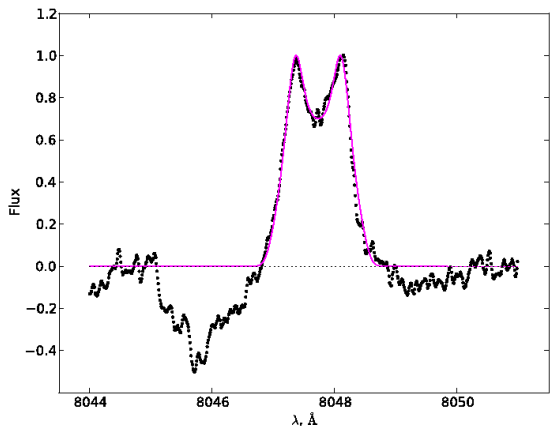}
 \caption{Mean, continuum-subtracted observed profile of an Fe emission line
in \iras\, \textit{(dots)} versus the best-fit model profile
from a Keplerian disc \textit{(solid)}.
Both profiles have been normalized to unity at maximum flux.
The broad absorption feature in the observed spectrum is due to the averaging
of a neighbouring photospheric line over the entire orbit.}
 \label{fig_Fe_Kepler}
 \end{figure}

In Figure \ref{fig_Fe_Kepler} we show another possibility to explain
the formation of the
double-peaked emission profiles -- in a Keplerian disc.
\citet{Smak1969} presented a simple kinematic model
of purely gaseous, optically thin disc of constant thickness
with density linearly dropping to zero at the outer edge.
Using this formulation, we could successfully fit Fe emission line profiles in \iras\,
with the following disc parameters:  
the ratio of the inner to outer radius of 0.15,
and the velocity at the outer edge of 12 km~s$^{\mathrm{-1}}$.  
Depending on the total mass of the system (0.9--1.6 M$_{\odot}$) and the inclination angle (30--70$\degr$),
this results in a disc with $R_{in}=0.2-1.3$ AU and $R_{out}=1.4-8.7$ AU.

Is this Fe disc circumcompanion or circumbinary? 
In Sec. \ref{sec_rv} we deduced that the semi-major axis of the pAGB primary
is 0.2 -- 0.4 AU and the secondary is likely more massive than the primary.
Hence, the distance between the companions does not exceed 0.4 -- 0.8 AU,
which is smaller than the outer radius of the Fe disc. 
This fact, together with the constancy of the RV of the emission lines,
imply that this disc is \textit{circumbinary}.
Furthermore, when applying a 2D radiative transfer code
\citep{Gielen2007,Gielen2009} to fit the SED of \iras, we obtain that the inner radius of
the circumbinary \textit{dusty} disc is 5 -- 10 AU.
Thus, the double-peaked Fe emission lines may signal the presence of a
gaseous Keplerian circumbinary disc, that is nestled within the sublimation boundary of the dusty disc.

In the prototypical disc object 89 Her the above discussed metal lines
form part of a much richer emission line spectrum  \citep{Climenhaga1987, Kipper2011}.
Based on the constancy of the RV, \citet{Waters1993} were first to propose
that emission could originate in a circumbinary disc.
Indeed, \citet{Bujarrabal2007} possibly detected such a disc
as an unresolved component in the interferometric maps of CO.
The existence of gas inside a dusty disc has been long anticipated
in the framework of the re-accretion hypothesis,
designed to explain a depletion pattern in some disc hosts.
The observational evidence of circumbinary gas, however, is largely missing.
CO studies normally probe material on a much larger scale (10$^3$--10$^5$ AU, \citealt{Bujarrabal2013}),
which could have been ejected in the preceding AGB stage.
In contrast, metal emission lines probe gas on AU scales,
and therefore provide a better insight into the current
mass loss/accretion.

Furthermore, the $\sim$10 times wider emission in H$\alpha$ may indicate
the presence of gas well inside the binary's orbit.
Giving that there is a jet emanating from the companion,
this hot gas could form in the \textit{circumcompanion} accretion disc.
Interestingly, the rare appearance of TiO in emission has been also associated
with the presence of an accretion disc.
Thus, \citet{Hillenbrand2012} proposed to explain TiO emission
in some young stellar objects (YSO) and Be stars by evaporating disc material at the base of the outflow,
where it is lifted up and exposed to the UV radiation from the accretion disc.
Besides \iras, TiO emission has been recently discovered
in a few candidate pAGB binaries in the Magellanic Clouds \citep{Wood2013}.

This raises some important questions about the type of the accretion (a wind, a Roche-lobe one,
or perhaps even accretion from the circumbinary disc),
and whether the three discs (the circumcompanion,
circumbinary, and the outer dust$+$gas one) could be possibly related to each other.
The exploration of these possibilities, however, is beyond the scope of this paper.

\subsection{Reflection of the circumbinary disc}\label{sec_refl}

What is the nature of the companion and could it explain the remaining peculiarities of the system:
the secondary set of spectral lines and the blueing of the colours in the light minima?

We applied the \textsc{FDBinary} code\footnote{http://sail.zpf.fer.hr/fdbinary} \citep{Ilijic2004}
to our spectral data set in the attempt to separate the two components.
The code performs separation of spectra in a spectroscopic double-lined binary star
in  Fourier space without the use of template spectra.
The input consists of the continuum-normalized spectra,
the relative fluxes of the two components ("light factors") per observation,
that  were taken from our double-gaussian decomposition of the CCFs,
and the first approximation for the orbital parameters.
In the output we obtained two identical sets of lines, whose shape and relative
depths depended somewhat on the
selected regions and the number of iterations, but in no case was there an indication
that the sets would correspond to two different spectral types.
This fact, along with the constancy of temperature in the opposite conjunctions,
implies that the companion must have an identical SpT and comparable luminosity to the primary.
And yet, only one star is seen in the light curve
(no substantial secondary minimum/maximum is observed over the RV period).
Also from the evolutionary point of view
it is very unlikely to find a system composed of two yellow evolved stars,
as pAGB stage is extremely short ($\sim$1000 years).

To resolve these difficulties, we propose an alternative explanation
for the secondary component of spectral lines,
in \iras, \bd, RR and potentially other inclined disc systems
with periodically distorted lines.
In this picture the companion is undetected, probably a non-evolved
dim MS star.
The secondary set of lines, which is most pronounced near the inferior conjunction of the primary,
would be the light of the primary that is reflected off the inner wall of the disc
and scattered into our line of sight.
The reflected image will move in the opposite direction
from the star and disappear during the superior conjunction of the latter,
explaining the CCF behaviour.

A very similar phenomenon of line splitting
due to the presence of reflected light was described for one classical
T Tauri binary KH 15D \citep{Herbst2008}.
Similar to \iras, KH 15D experiences periodic semi-regular deep declines,
that are explained by the orbital motion inside a precessing warped disc
\citep{Herbst2010, Windemuth2014}.
The SEDs of pre-main-sequence (pre-MS) objects are often indistinguishable from the pAGB ones,
pointing to the similarity of their discs \citep{Deruyter2006}.
One such property is the presence of grains
that are larger than in the interstellar medium.
Based on the comprehensive modelling,
\citet{Herbst2008} concluded that the reflected light in KH 15D provides the best evidence
that the particles in the disc grew to at least 1 mm in size. For pAGB discs the same was
previously inferred from the black-body slope of the far-IR excess,
which can only be measured for brighter sources.
Modelling of the scattered line profiles may prove to be a useful novel tool for studying
composition and kinematics of post-AGB discs \citep[e.g.][]{Grinin2006}.  
While in YSOs the particle growth is the first step toward the planet formation process,
the evolution of grains in post-AGB discs is unknown.

Large particles, however, produce gray scattering and settle to the
mid-plane of the disc, and hence can not explain the blueing of the
system in the minimum light when the primary sinks behind the disc
edge.  No trace of a hot companion is observed in our spectra either.
We propose that the blueing is due to the scattering by the inner wall
of the disc.  Since we do not resolve the system, the total flux in
the telescope beam is always a sum of direct and scattered light in
our line of sight.  During the inferior conjunction of the primary
(phase of minimum light) the amount of bluer, back-scattered (by the
farthest side of the inner disc wall) light may exceed the amount of
direct, reddened light from the primary, which is possible because,
unlike us, the inner disk wall always sees the star unobscured.  As a
result, in this phase the system will appear bluer to us than in the
unobscured phases.

The colours and the level of the scattered light bring stringent constraints on the size distribution and on the
chemo-physical properties of the dust grains, as well as on geometry of
the inner dusty disc. The reproduction of these observables will need
detailed radiative transfer models in which the angle and colour-dependent scattering,
as well as dust settling need to be incorporated. This is outside
the scope of this paper, however, and will be the subject of a
subsequent analysis.
The high level of optical scattering seems to be a common property of
the circumbinary discs, as also in 89 Her this has been detected using
optical interferometry. \cite{Hillen2013, Hillen2014} spatially resolved the
optical scattering component and deduced
that as much as 40\% of the optical flux of 89 Her is due to scattered
light.

\subsection{Circumstellar geometry}

\begin{figure*}
 \includegraphics[width=14cm]{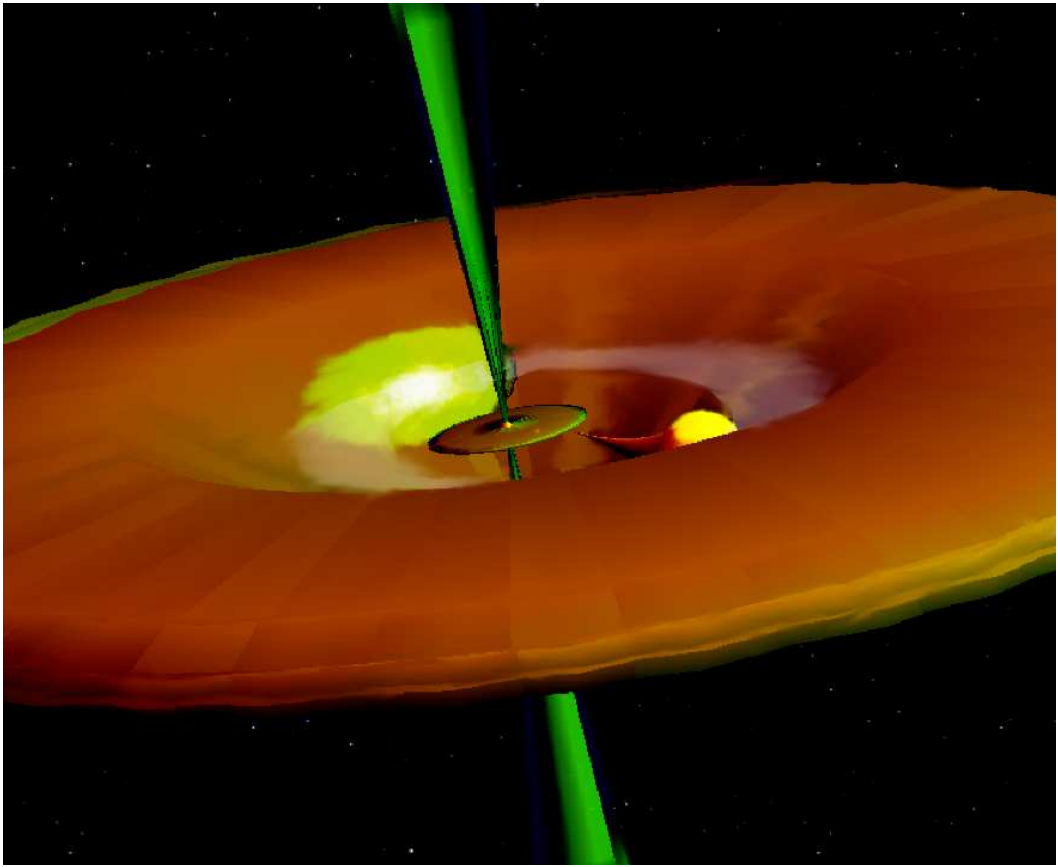}
 \caption{Artist's impression of \iras\, interacting binary system, depicted in the phase
 near light minimum. The pAGB primary is about to sink behind the dusty disc edge, giving way to
 the spot of reflected light on the opposite side of the circumbinary disc.}
 \label{fig_artist}
 \end{figure*}

In Fig. \ref{fig_artist} we present an artist's impression of the circumstellar environment of \iras\,
based on the discussion in the previous subsections.
Like in \bd, in this system the giant primary is transferring mass
to a, likely unevolved, companion, which results in the production of a pair of jets emanating from the secondary.
The system is surrounded by a dusty disc, and possibly also a smaller gaseous disc.
The new element in this picture is the high level of optical reflection 
from the primary on the wall of the circumbinary disc,
that mimics a twin companion in the spectra. This reflected light is
observed with a different Dopplershift and hence can be
differentiated from the component of direct light. 
The reflected spectrum and the
light amplitude are more pronounced in \iras\, than in \bd\, likely due to the higher inclination of the former.
The contribution of this reflected light is maximal at minimum light,
which indicates an efficient back-scattering and makes the system to
appear bluer when fainter.

\section{Conclusions}

We presented contemporaneous photometric and spectroscopic
observations of \iras, an SRd variable with an IR excess.
It is often assumed that variability of such stars is due to pulsations,
but there is no theoretical study yet that would consistently reproduce
all types of semi-regular variables.
We showed that \iras\, is a binary system with a circumbinary dusty disc.
These two properties, taken together, can explain the observed light variations
with a period of 127 days by an obscuration of the primary star
by the inner edge of the disc during inferior conjunctions. Given the
very smooth Kepler lightcurve, the photosphere of \iras\, is stable
and does not show any pulsations.
 
\iras\, presents a typical case of a class of pAGB binaries
where circumstellar matter is confined to a disc:
it is a low-gravity object, a single-line spectroscopic binary,
shows a depletion pattern of the refractory elements (albeit on a weaker side of the range),
and has a jet-like outflow from the companion,
based on the specific H$\alpha$ profiles.
The commonality of the latter phenomenon was realized 
thanks to the HERMES survey \citep{Gorlova2013}.
Furthermore, we detected static emission lines of Fe\,\textsc{i} and TiO
that are most pronounced in the spectra taken near minimum light.
We successfully fit the double-peaked Fe lines with a circumbinary Keplerian disc,
that may be the first evidence of the star-disc interaction leading
to the depletion pattern in pAGB atmospheres.
The nature of the invisible companion remains unknown.
We detected a secondary set of spectral lines, but prove that
it must be a reflected spectrum of the primary on the disc grains,
rather than a spectrum of a physical companion. 

The single case of \iras\, presented here can not
rule out pulsations for all SRd variables.
What our study illustrates is that the presence of a circumbinary
dusty disc
may have profound effects on the light and RV curves, and therefore needs to
be considered on a par with the more traditional factors,
such as stellar eclipses and pulsations.

\section*{Acknowledgments}

We would like to thank Drs. P. Harmanec and H. Hensberge for the fruitful
discussions, the numerous observers from the HERMES consortium institutions
for collecting the spectral data, and the anonymous referee for the comments that helped to sharpen the text.
This study is based on observations made with the Mercator Telescope,
operated on the island of La Palma by the Flemish Community,
at the Spanish Observatorio del Roque de los Muchachos of the Instituto de Astrofisica de Canarias;
and with the HERMES spectrograph,
which is supported by the Fund for Scientific Research of Flanders (FWO), Belgium,
the Research Council of K.U.Leuven, Belgium, the Fonds National de la Recherche Scientifique (F.R.S.-FNRS), Belgium,
the Royal Observatory of Belgium, the Observatoire de Geneve, Switzerland and the Th\"uringer Landessternwarte Tautenburg, Germany. 


\bibliographystyle{mn2e}
\bibliography{ngorlova}


\appendix

\onecolumn
\section[]{New measurements of {\textit{IRAS}~19135$+$3937}}

\begin{center}
\begin{longtable}{@{}cccccc}
\caption{$UBV$ photometry in 2012-2013}\\
\label{tab_ubv} \\ 
\hline
JD&$V$&$B$&$U$&$B-V$&$U-B$\\
\hline 
\endfirsthead
\multicolumn{6}{c}%
 {\tablename\ \thetable\ -- \textit{Continued from previous page}} \\ \hline 
  JD&$V$&$B$&$U$&$B-V$&$U-B$\\
 \hline \\
\endhead
\multicolumn{6}{r}{\textit{Continued on next page}} \\ \hline
\endfoot
\hline
\endlastfoot
2456093.447&	11.730&	12.231&	12.521&	0.501&	0.290\\
2456094.482&	11.716&	12.236&	12.485&	0.520&	0.249\\
2456095.375&	11.731&	12.199&	12.522&	0.468&	0.323\\
2456101.456&	11.674&	12.158&	12.454&	0.484&	0.296\\
2456119.425&	11.299&	11.854&	12.194&	0.555&	0.340\\
2456121.473&	11.204&	11.781&	12.147&	0.577&	0.366\\
2456122.437&	11.180&	11.752&	12.086&	0.572&	0.334\\
2456128.378&	10.955&	11.550&	11.940& 0.595&	0.390\\
2456131.353&	10.859&	11.454&	11.874&	0.595&	0.420\\
2456133.355&	10.781&	11.402&	11.854&	0.621&	0.452\\
2456136.312&	10.719&	11.344&	11.773&	0.625&	0.429\\
2456137.447&	10.702&	11.312&	11.758&	0.610&	0.446\\
2456146.374&	10.615&	11.263&	11.732&	0.648&	0.469\\
2456147.382&	10.628&	11.286&	11.743&	0.658&	0.457\\
2456151.353&	10.644&	11.295&	11.765&	0.651&	0.470\\
2456159.404&	10.793&	11.448&	11.895&	0.655&	0.447\\
2456161.360&	10.844&	11.488&	11.951&	0.644&	0.463\\
2456166.472&	10.996&	11.600&	12.012&	0.604&	0.412\\
2456176.375&	11.318&	11.881&	12.234&	0.563&	0.353\\
2456177.302&	11.372&	11.913&	12.269&	0.541&	0.356\\
2456186.302&	11.597&	12.103&	12.365&	0.506&	0.262\\
2456189.333&	11.598&	12.094&	12.382&	0.496&	0.288\\
2456200.264&	11.734&	12.215&	12.504&	0.481&	0.289\\
2456202.319&	11.740&	12.233&	12.545&	0.493&	0.312\\
2456208.313&	11.722&	12.210&	12.487&	0.488&	0.277\\
2456216.232&	11.687&	12.180&	12.504&	0.493&	0.324\\
2456224.236&	11.695&	12.163&	12.481&	0.468&	0.318\\
2456241.222&	11.352&	11.878&	12.192&	0.526&	0.314\\
2456249.208&	11.017&	11.591&	11.964&	0.574&	0.373\\
2456270.188&	10.613&	11.249&	11.660&	0.636&	0.411\\
2456276.184&	10.674&	11.318&	11.760&	0.644&	0.442\\
2456405.565&	10.669&	11.349&	11.831&	0.680&	0.482\\
2456406.512&	10.682&	11.352&	11.852&	0.670&	0.500\\
2456420.503&	10.908&	11.577&	12.045&	0.669&	0.468\\
2456422.506&	10.945&	11.627&	12.059&	0.682&	0.432\\
2456431.517&	11.229&	11.865&	12.216&	0.636&	0.351\\
2456434.490&	11.315&	11.927&	12.278&	0.612&	0.351\\
2456445.479&	11.641&	12.184&	12.455&	0.543&	0.271\\
2456454.514&	11.743&	12.239&	12.577&	0.496&	0.338\\
2456463.510&	11.760&	12.265&	12.546&	0.505&	0.281\\
2456472.379&	11.748&	12.231&	12.528&	0.483&	0.297\\
2456473.392&	11.725&	12.240&	12.547&	0.515&	0.307\\
2456479.367&	11.676&	12.171&	12.510&	0.495&	0.339\\
2456482.428&	11.609&	12.113&	12.434&	0.504&	0.321\\
2456484.401&	11.568&	12.102&	12.409&	0.534&	0.307\\
2456485.356&	11.558&	12.072&	12.402&	0.514&	0.330\\
2456487.401&	11.504&	12.042&	12.324&	0.538&	0.282\\
2456489.405&	11.437&	11.988&	12.332&	0.551&	0.344\\
2456492.399&	11.343&	11.914&	12.309&	0.571&	0.395\\
2456504.440&	10.860&	11.506&	11.933&	0.646&	0.427\\
2456506.431&	10.801&	11.437&	11.891&	0.636&	0.454\\
2456510.426&	10.679&	11.347&	11.801&	0.668&	0.454\\
2456513.336&	10.649&	11.306&	11.773&	0.657&	0.467\\
2456513.413&	10.637&	11.307&	11.737&	0.670&	0.430\\
2456514.339&	10.627&	11.284&	11.787&	0.657&	0.503\\
2456515.397&	10.606&	11.279&	11.744&	0.673&	0.465\\
2456517.412&	10.594&	11.244&	11.751&	0.650&	0.507\\
2456518.358&	10.592&	11.258&	11.755&	0.666&	0.497\\
2456519.424&	10.572&	11.248&	11.729&	0.676&	0.481\\
2456520.379&	10.575&	11.249&	11.697&	0.674&	0.448\\
2456531.380&	10.635&	11.304&	11.776&	0.669&	0.472\\
2456533.410&	10.670&	11.336&	11.819&	0.666&	0.483\\
2456545.372&	10.965&	11.621&	12.043&	0.656&	0.422\\
2456563.302&	11.615&	12.127&	12.454&	0.512&	0.327\\
2456573.270&	11.749&	12.248&	12.519&	0.499&	0.271\\
2456574.326&	11.738&	12.240&	12.508&	0.502&	0.268\\
2456577.271&	11.781&	12.280&	12.602&	0.499&	0.322\\
2456586.236&	11.807&	12.324&	12.649&	0.517&	0.325\\
2456591.264&	11.790&	12.304&	12.603&	0.514&	0.299\\
2456597.284&	11.731&	12.222&	12.575&	0.491&	0.353\\
2456602.292&	11.702&	12.219&	12.505&	0.517&	0.286\\
2456606.257&	11.687&	12.183&	12.477&	0.496&	0.294\\
2456607.236&	11.695&	12.203&	12.510&	0.508&	0.307\\
2456612.229&	11.688&	12.176&	12.463&	0.488&	0.287\\
\end{longtable}
\end{center}


\begin{table}
\centering
\caption{Radial velocities in 2009-2013}\label{tab_rv}
\begin{tabular}{crr}
 \hline

 JD & $RV_{main}$ & $RV_{sec}$ \\
 (d) & (km~s$^{\mathrm{-1}}$) & (km~s$^{\mathrm{-1}}$) \\
 \hline

2454993.45906 & 13.90  & -20.90 \\
2455003.50920 & 3.29   & -22.24 \\
2455003.52830 & 3.53   & -22.56 \\
2455003.54740 & 3.62   & -19.50 \\
2455012.40789 & -6.29  & 37.08 \\
2455012.42723 & -6.05  & 33.09 \\
2455012.44660 & -6.23  & 34.26 \\
2455025.56787 & -16.78 & 25.22 \\
2455034.55947 & -18.74 & 22.30 \\
2455061.44461 & -1.69  & N/A \\
2455084.42500 & 17.32  & -7.19 \\
2455423.61170 & -16.83 & 14.74 \\
2455430.45750 & -12.33 & 12.36 \\
2455501.33556 & 13.52  & -21.61 \\
2455507.32911 & 7.97   & -18.11 \\
2455650.71184 & -7.63  & 33.15 \\
2455707.66632 & 5.21   & -10.86 \\
2455714.50939 & 6.58   & -13.75 \\
2455745.56824 & 19.68  & -22.63 \\
2455761.54727 & 7.06   & -24.93 \\
2455763.56835 & 4.86   & -24.68 \\
2455773.61864 & -4.93  & 30.83 \\
2455778.54334 & -8.67  & 29.72 \\
2455784.49019 & -12.61 & 29.06 \\
2455807.49337 & -17.13 & 10.34 \\
2455828.42879 & 0.22   & N/A \\
2455837.44068 & -1.57  & N/A \\
2455843.35922 & 2.57   & -16.91 \\
2455873.34316 & 20.68  & -17.48 \\
2455993.73419 & 19.49  & -20.90 \\
2456010.70972 & 13.30  & -20.14 \\
2456015.64342 & 7.94   & -24.72 \\
2456033.67080 & -9.28  & 27.91 \\
2456056.54496 & -16.41 & 14.72 \\
2456082.59648 & 0.01   & N/A \\
2456087.52359 & -1.19  & N/A \\
2456088.53941 & -1.90  & N/A \\
2456103.40544 & 11.66  & -14.23 \\
2456107.43456 & 11.77  & -18.93 \\
2456117.48552 & 21.29  & -18.95 \\
2456127.54549 & 20.26  & -19.66 \\
2456136.41771 & 14.06  & -22.11 \\
2456140.50515 & 9.14   & -24.19 \\
2456144.61982 & 4.28   & N/A \\
2456146.48406 & 3.85   & N/A \\
2456152.45858 & -2.36  & 30.74 \\
2456157.49963 & -7.44  & 29.25 \\
2456165.48663 & -13.20 & 27.11 \\
2456176.53346 & -14.96 & 22.14 \\
2456178.58460 & -13.87 & 21.02 \\
2456188.51469 & -9.01  & 16.36 \\
2456194.44581 & -8.25  & N/A \\
2456392.74729 & 12.65  & -22.95 \\
2456441.69257 & -17.15 & 11.11 \\
2456457.45702 & 1.96   & N/A \\
2456468.48879 & 2.90   & -14.97 \\
2456477.43936 & 2.73   & -15.64 \\
2456488.56607 & 17.36  & -15.94 \\
2456512.62342 & 18.19  & -22.41 \\
2456544.46648 & -11.58 & 26.57 \\
2456566.49203 & -11.40 & 16.88 \\

 \hline
\end{tabular}

HERMES radial velocities for the main and secondary
components in the cross-correlation function of \iras,
as described in Sec. \ref{sec_rv}.
``N/A'' stands for 'not available' and designates phases where it was not possible
to disentangle the secondary component.

\label{lastpage}

\end{table}

\bsp

\end{document}